\documentclass[lettersize,journal]{IEEEtran}
\usepackage{amsmath,amsfonts}
\usepackage{algorithmic}
\usepackage{algorithm}
\usepackage{array}
\usepackage[caption=false,font=normalsize,labelfont=sf,textfont=sf]{subfig}
\usepackage{textcomp}
\usepackage{stfloats}
\usepackage{url}
\usepackage{bm}
\usepackage{verbatim}
\usepackage{graphicx}
\usepackage{diagbox}
\usepackage{booktabs}
\usepackage{cite}
\begin{document}

\title{Maximum Correntropy Value Decomposition for Multi-agent Deep Reinforcemen Learning}

\author{Kai Liu, Tianxian Zhang, Lingjiang Kong
        }



\maketitle

\begin{abstract}
We explore value decomposition solutions for multi-agent deep reinforcement learning in the popular paradigm of centralized training with decentralized execution(CTDE). 
As the recognized best solution to CTDE, Weighted QMIX is cutting-edge on StarCraft Multi-agent Challenge (SMAC), with a weighting scheme implemented on QMIX to place more emphasis on the optimal joint actions. 
However, the fixed weight requires manual tuning according to the application scenarios, which painfully prevents Weighted QMIX from being used in broader engineering applications. 
In this paper, we first demonstrate the flaw of Weighted QMIX using an ordinary One-Step Matrix Game (OMG), that no matter how the weight is chosen, Weighted QMIX struggles to deal with non-monotonic value decomposition problems with a large variance of reward distributions.
Then we characterize the problem of value decomposition as an Underfitting One-edged Robust Regression problem and make the first attempt to give a solution to the value decomposition problem from the perspective of information-theoretical learning.
We introduce the Maximum Correntropy Criterion (MCC) as a cost function to dynamically adapt the weight to eliminate the effects of minimum in reward distributions. 
We simplify the implementation and propose a new algorithm called MCVD. 
A preliminary experiment conducted on OMG shows that MCVD could deal with non-monotonic value decomposition problems with a large tolerance of kernel bandwidth selection. 
Further experiments are carried out on Cooperative-Navigation and multiple SMAC scenarios, where MCVD exhibits unprecedented ease of implementation, broad applicability, and stability.
\end{abstract}


\section{Introduction}
\IEEEPARstart{T}{he} decentralized partially observable Markov decision process (Dec-POMDP)\cite{oliehoek2016concise} is one of the most universal and acknowledged models in multi-agent sequential decision-making tasks, where all the agents share a collective reward in cooperative settings. 
Although the model is easy for human imagination and engineering application, it makes algorithm design much more difficult\cite{busoniu2008comprehensive, hernandez2019survey, nguyen2020deep}.
To learn good behaviors in Dec-POMDP, each agent needs to extract appropriate individual reward from the collective reward, resulting in one of the most widely studied areas of the multi-agent system: Credit assignment\cite{foerster2018counterfactual, iqbal2019actor}.
In value-based (VB) multi-agent reinforcement learning(MARL), the collective reward is linearly linked to the joint action-value by the Bellman Equation, and the credit assignment problem is naturally transformed to joint action-value decomposition\cite{hauskrecht2000value,  liu2022self} (also called value function factorization or value decomposition in brief in many papers).

Various algorithms have been proposed to address this challenge in recent years. 
VDN\cite{VDN} did the first attempt, which factorize joint action-values  into a form of the sum of individual action values. 
QMIX\cite{qmix-j} extends the expression of additive value factorization to a monotonic form with the help of the hyper-network\cite{ha2016hypernetworks}, thereby covering a richer variety of MARL problems than VDN.
QTRAN\cite{son2019qtran} further extends the expression into a non-monotonic form, differentiated learning is achieved by calculating the loss of optimal joint actions and non-optimal joint actions respectively.
Despite well-researched theories, QTRAN does not perform well in some routine tests, such as SMAC\cite{samvelyan2019starcraft}.
IGM (Individual Global Max) principle is a definition proposed in QTRAN, which emphasizes the equivalence requirement between optimal joint actions and individual optimal actions, and relaxes the equivalence requirement between non-optimal joint actions and individual non-optimal actions, thus greatly reducing the learning cost of value decomposition problems.
Under the guidance of IGM, researchers attempt to construct the combinations of individual action values ${Q_{jt}}$ as a hypersurface overlying the joint action-values $y$: ${Q_{jt}} \ge y$.
Weighted QMIX\cite{rashid2020weighted} borrows inspiration from IGM, and introduced a weighting function to place more emphasis on approximating the optimal joint actions. 
With elaborately designed weight, to our best knowledge, Weighted QMIX (especially Optimistically-Weighted QMIX, OW-QMIX) achieves the best results among various non-monotonic value decomposition algorithms\cite{son2019qtran,wang2020qplex,iqbal2019actor}.

However, results obtained by Weighted QMIX greedy rely on the manually selected weights, and the optimal weights are not the same for different scenarios. 
Weighted QMIX does not give guidance on the selection of weights, merely tests multiple weights and selects the best one for exhibition, where $\alpha = 0.1$ in Predator-Prey and $\alpha = 0.5$ in SMAC. 
The effect of varying $\alpha$ in SMAC was conducted but not in Predator Prey. 
Essentially, Weighted QMIX works like a system consisting of two types of elastic bands with different tension coefficients, when ${Q_{jt}} < y$, the coefficient $w$ is 1, otherwise, $w= \alpha  < 1$, and the band will never fail. 
Using this metaphor, we all know that even with a very small tension coefficient, if the distance is large enough, we can still get considerable tensile force. 
The same is true for Weighted QMIX, as long as the TD (temporal-difference)\cite{cai2019td} error is large enough, a large loss will be obtained.
Since the weights are fixed scalars, Weighted QMIX does not change the fact that the gradients of all joint action-values have a positive correlation with their TD errors, nor does it discriminate the weight between optimal joint actions and non-optimal ones. 
Therefore, it cannot solve the non-monotonic value decomposition problem in essence.

We experimentally verify the above assertion using an ordinary OMG and find that Weighted QMIX struggles to decompose non-monotonic joint rewards with large variances.
The fundamental reason is that the selection of an appropriate fixed weight for Weighted QMIX depends heavily on the distribution of samples,
which severely weakens the scalability of Weighted QMIX.
To our best knowledge, the above problem is not yet available in the open literature.
And designing a dynamic weight instead of manual tuning is indeed the main pursuit of this paper.

Through analyzing the drawbacks of Weighted QMIX, we rethink the value decomposition problem and characterize it as an Underfitting One-edged Robust Regression problem.
Under this qualitative analysis, we make the first attempt to give a solution to the value decomposition problem from the perspective of information-theoretic learning.
An information theory concept, maximum correntropy criterion\cite{principe2010entropy}, is introduced into the field of value decomposition.
MCC has been widely used in Adaptive Filter and Kalman Filter in recent years\cite{chen2014steady,mckf, d-mckf}, which achieves good performances in eliminating outlier noise through Gaussian kernel function. 
Compared with the well-known minimum mean square error (MMSE) criterion, MCC adds a negative exponential coefficient to the original gradient, thus greatly weakening the effect of large errors. 
Using the above mechanism, we propose the Maximum Correntropy Value Decomposition (MCVD) algorithm for multi-agent deep reinforcement learning, where minimum rewards are treated as outliers and their interference in selecting optimal action values is eliminated.

We discuss and analyze the pros and cons of networks used by existing MARL algorithms, and form the overall framework of MCVD.
Compared with existing algorithms, the implementation of MCVD is extremely simple and does not need parameter tunes according to specific environments. 
Extensive experiments are conducted and show that MCVD has stable and excellent performances in both monotonic and non-monotonic scenarios.
The broader applicability and stability of MCVD make it promising for widely used in engineering.

The rest of this paper is organized as follows: 
In Section 2, the relevant background is introduced.
The flaw of Weighted QMIX is demonstrated in Section 3.
In Section 4, we provide a new solution for value decomposition from the perspective of information-theoretical learning and propose a new algorithm called MCVD.
And in Section 5, experiments are conducted in three environments to demonstrate the superiority of MCVD.
Finally, Section 6 concludes the paper.

\section{Background}
\subsection{Dec-POMDP}
The cooperative multi-agent tasks are always formulated as Dec-POMDP, which is usually defined as a tuple $({{\cal N}},{{\cal S}},\bm{{\cal O}},\bm{{\cal A}},P,{{\cal R}},\gamma )$, where ${{\cal N}} = {1,...,i,...,N}$ is a finite index set of agents, ${{\cal S}}$ is the set of environmental states, representing the global configurations generated from the environment, which can not be obtained by agents during execution.
We denote $\bm{{\cal O}} = {{{\cal O}}_1} \times {{{\cal O}}_2} \times  \cdots  \times {{{\cal O}}_n}$\footnote{We use the subscript of $({ \cdot _i})$ to distinguish between agent $i$ and other cooperators.}
, where ${{{\cal O}}_i}$ is the set of individual observations of agent $i$; and $\bm{{\cal A}} = {{{\cal A}}_1} \times {{{\cal A}}_2} \times  \cdots  \times {{{\cal A}}_n}$, where ${{{\cal A}}_i}$ is the set of actions of agent $i$. 
In the phase of sample generation, each agent receives an individual observation from the environment according to the state: ${{\cal S}} \mapsto {{{\cal O}}_i}$ and select action ${{{\cal A}}_i}$ using a stochastic initialized individual policy ${{\bf{\pi }}_{{i}}}$. The environment generates the next state and collective reward based on the current state, joint actions and state transition function, which can be formulated as: $({\cal R},{\cal S'}) = P(\bm{{\cal A}}|\cal S)$. The objective of joint policy ${\bf{\pi }} = {{\bf{\pi }}_{{1}}}\times{{\bf{\pi }}_{{2}}}\times...\times{{\bf{\pi }}_{{n}}}$ is to maximize the expected discounted return (cumulative discounted future reward) $R = \mathbb{E}[\sum\limits_{j = 1}^\infty  {\gamma^j {r}],r \in \cal R}$, and $\gamma  \in [0,1]$ is the discount factor.

\subsection{Deep Q-Learning}
\label{sec:DQN}
Deep Q-learning (DQN)\footnote{The DQN described in this section is in the form of Double DQN (DDQN)\cite{DDQN}, DDQN is so commonly used that most studies do not strictly distinguish it from DQN.} is the most widely used algorithm in reinforcement learning and also a representation of value-based (VB) method\cite{dqn-atari}, compared with most policy gradient (PG) method\cite{schulman2015trust, ppo}, DQN uses a replay buffer ${{\cal B}}$ to store the transition tuple $ < s,a,r,s' > $, greatly improves the sample utilization and algorithm stability. The action-value function in DQN is parameterized by $\theta $, which is learned by sampling transitions from the replay buffer and minimizing the mean square TD errors:
\begin{equation}
\label{Eq:DQN}
   {\cal L}(\theta ) = {\mathbb{E}_{s,a \sim b}} {[{{(y - Q(s,a;\theta ))}^2}]},
\end{equation}
where $y = r + \gamma (1-t)Q'(s',\bm{\bar u}';\theta '){|_{\bm{\bar u}' = \mathop {\max }\limits_u Q'(s;\theta ')}}$ is the greedy Bellman Equation\footnote{We denote $a$ as the indices of non-gradient action values sampled from the replay buffer and $u$ as the indices of gradient action values calculated in real time during training.}, $b$ is a batch sampled from the replay buffer, $t$ is the terminate flag, ${\theta '}$ are the parameters of the target network that are periodically updated by copying from ${\theta}$ and keep constant for a period of iterations.

\subsection{Joint Value Function and CTDE}
In the setting of value-based MARL, each agent has an independent or shared action-value network\footnote{If a shared value network is used, the input vector will be encoded according to the index of agent.} called individual action-value network. The above network takes observation $o_i$ as input\footnote{If RNN network such as LSTM or GRU is used, the historical information ${\tau _i}$ is also required.} and outputs individual action values $Q_i(o_i,\cdot)$ (if a deterministic action is selected, $Q_i(o_i,a_i)$).
Since Dec-POMDP is set up with only collective reward $r$, the concept of joint action values ${Q_{jt}}(\bm{o},\bm{a})$ needs to be stylized to implement the greedy Bellman Equation in Sec:\ref{sec:DQN}. 
\begin{equation}
\begin{array}{l}
{{{\cal L}}_{td}(\theta)} = {\mathbb{E}_{\bm{o},\bm{a} \sim b}} {[(y - {Q_{jt}}(\bm{o},\bm{a};\theta))^2]},\\
y = r + \gamma (1-t){{Q'}_{jt}}(\bm{o}',\bm{\bar u}';\theta '){|_{\bm{\bar u}' = \mathop {\arg \max }\limits_{\bm{u}} {{Q'}_{jt}}(\bm{o}', \cdot ;\theta')}}
\end{array}
\label{joint value function}
\end{equation}
Then follows the questions of what kind of relationship should be maintained between ${Q_{jt}}(\bm{o},\bm{a};\theta)$ and $Q_i(o_i,a_i;\phi _ i)$, which is known as the value decomposition problem, and the scalability of ${\bm{\bar u}' = \mathop {\arg \max }\limits_{\bm{u}} {{Q'}_{jt}}(\bm{o}', \cdot ;\theta ')}$.

Centralized training with decentralized execution\cite{oliehoek2008optimal,kraemer2016multi,RQN} is a popular paradigm of cooperative MARL.
In the centralized training process, agents are granted access to other agents' observations $o^{-i}$ or the global states $s$. 
However, during the decentralized execution process, each agent makes decision based on its partial observation $o_i$. 
This imitates the partial observability and communication constraints of individuals in engineering.
With the help of CTDE, $Q_{jt}$ can achieve complex function mapping with joint action combination $\bm{a} = {a_1} \times {a_2} \times  \cdots  \times {a_N}$ as well as state $s$ through various networks, usually denoted as ${Q_{jt}}(s,\bm{a};\theta)$ instead of ${Q}_{jt}(\bm{o}, \bm{a} ;\theta)$\footnote{In order to simplify the writing, the parameters of the neural network will be omitted later.}.

\subsection{Value Decomposition and IGM}

VDN is the pioneering work of value decomposition, it represent ${Q_{jt}}$ as the sum of individual action values ${\{ {Q_i}\} _{i \in {{\cal N}}}}$, which can be formulated as:
\begin{equation}
    {Q_{jt}}(s,\bm{a}) = \sum\limits_{i = 1}^N {{Q_i}({o_i},{a_i}),} 
\end{equation}
\begin{equation}
    or\quad \frac{{\partial {Q_{jt}}(s,\bm{a})}}{{\partial {Q_i}({o_i},{a_i})}} \equiv 1,\forall i \in \cal N.
\end{equation}

QMIX is another landmark of value decomposition solutions, which extends the expressiveness of joint action-value function. QMIX enforce a monotonic constraint on the relationship between ${Q_{jt}}$ and ${\{ {Q_i}\} _{i \in {{\cal N}}}}$:
\begin{equation}
    \quad \frac{{\partial {Q_{jt}}(s,\bm{a})}}{{\partial {Q_i}({o_i},{a_i})}} \ge 0,\quad \forall i \in \cal N.
\end{equation}
The monotonicity is achieved through a novel hyper-network\cite{ha2016hypernetworks}, which takes the outputs of the branch neural network as the parameters of the trunk neural network. 

Kyunghwan Son and Daewoo Kim point out that both VDN and QMIX could only deal with a small variety of MARL problems, they construct a non-monotonic joint value function through OMG and unveil the limitations of VDN and QMIX that if ${Q_{jt}}(s,\bm{a})$ wants to guarantee an exact relationship with the value each individual action, such as summation or monotonicity, it can never establish an equality relation with $y$ under all action combinations, i.e. $\exists \bm{a} \in \cal A$:
\begin{equation}
{Q_{jt}}(s,\bm{a}) \equiv \left( {\begin{array}{*{20}{c}}
{{Q_1}({o_1},{a_1})}\\
 \cdots \\
{{Q_N}({o_N},{a_N})}
\end{array}} \right) \ne y,
\end{equation}

Their solution for value decomposition is QTRAN, a CTDE method based on inequality relationships that guarantee more general factorization than VDN and QMIX.
\begin{equation}
\begin{array}{l}
{Q_{jt}}(s,\bm{a})  - {\hat Q_{jt}}(s,\bm{a}) + {V_{jt}}(s) = \left\{ {\begin{array}{*{20}{l}}
{0\quad \bm{a} = \bm{\bar a},}\\
{ \ge 0\quad \bm{a} \ne \bm{\bar a},}
\end{array}} \right.\\
{V_{jt}}(s) = \mathop {\max }\limits_{\bm{a}} {\hat Q_{jt}}(s,\bm{a}) - {Q_{jt}}(s,\bm{a}) .
\end{array}
\end{equation}
Where ${Q_{jt}}(s,\bm{a}) = \sum\limits_{i = 1}^N {{Q_i}({o_i},{a_i})}$, and ${\hat{Q}_{jt}}(s,\bm{a})$  is the joint action-value approximation function:
${\hat{Q}_{jt}}(s,\bm{a})\approx y$\footnote{For a network with sufficient fitting ability, ${\hat{Q}_{jt}}(s,\bm{a})\approx y$, $y$ is a common expression in various studies, but it cannot accurately express the optimal joint action value. Readers can regard $y$ as a general expression, and ${\hat{Q}_{jt}}(s,\bm{a})$ is a convenient expression for indexing.}, which is mainly used for calculating the loss of optimal individual actions. 
$(\bar  \cdot )$ denotes the index of optimal action, for example, ${{\bar a}_i} = \mathop {\arg \max }\limits_{{a_i}} {Q_i}({o_i}, \cdot )$, and $\bm{\bar a} = {{\bar a}_1} \times {{\bar a}_2} \times  \cdots {{\bar a}_n}$.

IGM\cite{son2019qtran} is a generally accepted principle proposed in QTRAN, which asserts the consistency between joint action-value ${\hat{Q}_{jt}}(s,\bm{a})$ and individual action values ${\{ {Q_i}\} _{i \in {{\cal N}}}}$ when the local greedy actions were selected.
\begin{equation}
\mathop {\arg \max }\limits_{\bm{a}} {\hat{Q}_{jt}}(s,\bm{a}) = \left( {\begin{array}{*{20}{c}}
{\mathop {\arg \max }\limits_{{a_1}} {Q_1}({o_1},{a_1})}\\
{...}\\
{\mathop {\arg \max }\limits_{{a_N}} {Q_N}({o_N},{a_N})}
\end{array}} \right),
\end{equation}
Simply put, the optimal joint action-values should obtained if and only if each individual agent achieves the optimal action value. IGM only guarantees global and local optimal consistency, and there is no restriction on non-optimal actions. If IGM holds, we can say that ${\hat{Q}_{jt}}(s,\bm{a})$ is factorized by ${\{ {Q_i}\} _{i \in {{\cal N}}}}$ or that ${\{ {Q_i}\} _{i \in {{\cal N}}}}$ are factors of ${\hat{Q}_{jt}}(s,\bm{a})$.
Under the guidance of IGM, in recent studies, researchers have devoted themselves to constructing $Q_{jt}$ as a hypersurface overlying $y$: $Q_{jt} \ge y$, if and only if $\bm{a} = \bm{\bar a},{Q_{jt}} = y$, Weighted QMIX is also trying to achieve this goal.

In this section, we briefly introduced three cornerstone value decomposition methods related to this study. In the past two years, many multi-agent policy gradient methods\cite{yu2021mappo,gu2021matrpo,kuba2021trust,hu2021rethinking} have been proposed to handle the non-monotonic problem and achieved considerable results. However, since they don't have the individual or joint value functions, thus not in the same category of value decomposition as this paper, we will not elaborate on them, nor compare with them in future experiments.

\section{Weighted QMIX Operator}
In this section, we will prove the irrationality of weight design in Weighted QMIX through an ordinary One-Step Matrix Game with four different handcrafted weights, and explain the impracticability of weight selection in application from a theoretical perspective.

\subsection{Fundamental Operator of Weighted QMIX}
We identify with the fundamental operator of Weighted QMIX:
\begin{equation}
{{{\cal L}}_{td}} = {\mathbb{E}_{s,\bm{a} \sim b}} {[w(s,\bm{a}){(y - {Q_{jt}}(s,\bm{a}))^2}]},
\end{equation}
where the weighting function $w$ depends on the state and joint action, and $y$ is obtained iteratively from the Bellman equation, similar to Eq.\ref{joint value function}, which incorporates different approximation methods for calculating joint action-values.

\subsection{Implementation of Weighted QMIX}
\label{im wqmix}
The choice of weighting is vital to ensure that Weighted QMIX can overcome the boundaries of QMIX to solve non-monotonic problems.
To reach scalability, 
two weighting design methods have been proposed by Tabish Rashid et, al.
Take OW-QMIX as an example\footnote{The choice has two main reasons. On the one hand, OW-QMIX is more in line with the originalism of Weighted QMIX, which does not contain any special calculation of the temporary optimal individual action value. 
On the other hand, OW-QMIX has better performance than CW-QMIX in most of the experiments, and it is easier to illustrate graphically.},
the implementation of the weighting function is:
\begin{equation}
w(s,\bm{a}) = \left\{ {\begin{array}{*{20}{l}}
1\\
\alpha
\end{array}} 
\quad \right.\begin{array}{*{20}{l}}
{{Q_{jt}}(s,\bm{a}) < y,}\\
{otherwise.}
\end{array}
\label{eq:ow-qmix}
\end{equation}

\begin{figure}[!t]
    \centering
    \includegraphics[width = 3in]{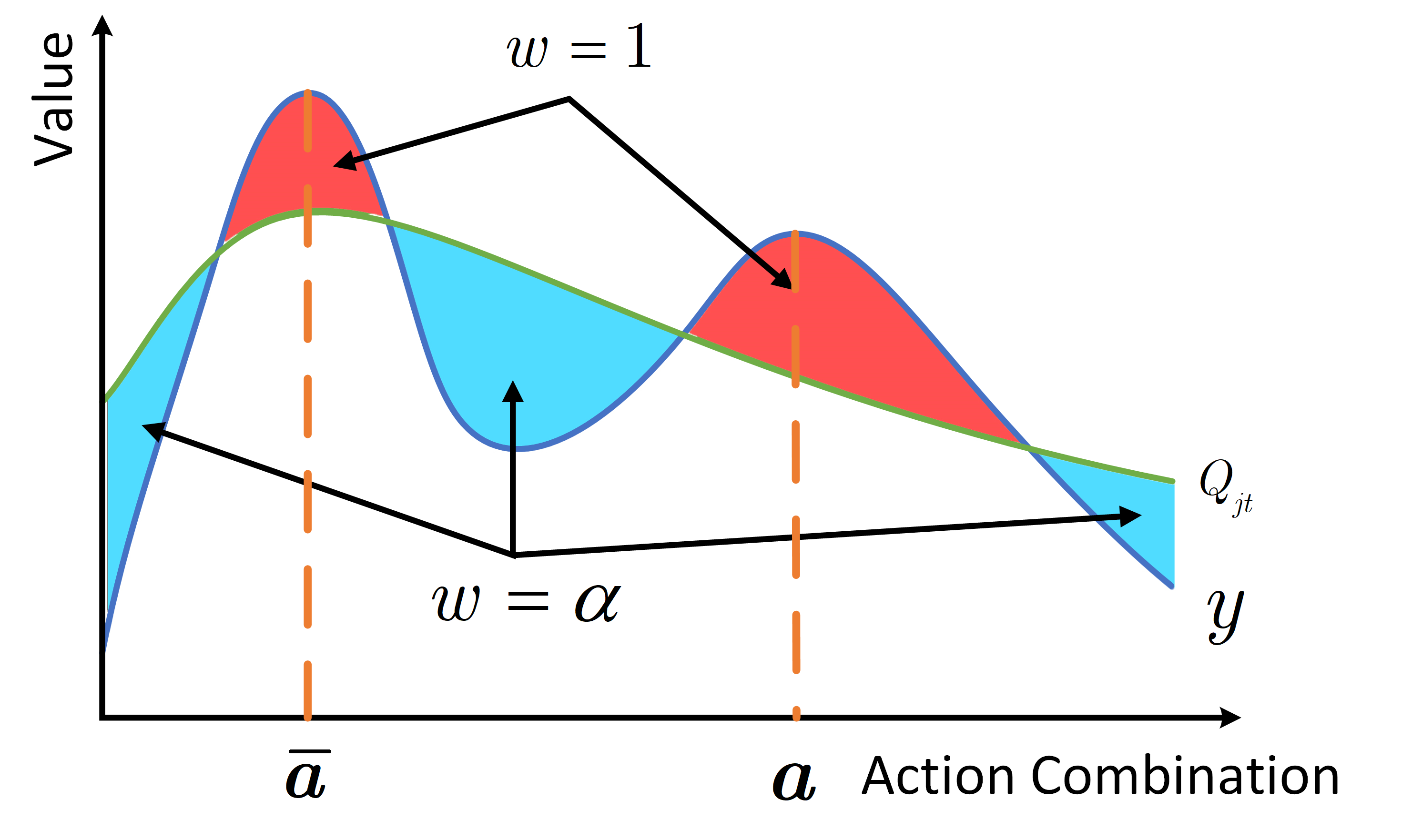}
    \caption{Illustration of OW-QMIX. The x-axis represents the sample of individual action combinations, the red region represents that ${{Q_{jt}}(s,\bm{a}) < y}$ where weight $w$ is equal to 1, the blue area represents that ${{Q_{jt}}(s,\bm{a}) > y}$, where $w$ is equal to $\alpha$, and $\alpha < 1$. }
    \label{fig:ow-qmix}
\end{figure}

OW-QMIX works like a system consisting of two types of elastic bands with different tension coefficients, which can be graphically illustrated by Fig. \ref{fig:ow-qmix}

Since there is a large number of agents, and $Q_{jt}$ is the output of the hyper-network, i.e. the monotonic combination of individual action value of all agents. 
Thus if a high-dimensional $Q_{jt}$ is squeezed in two dimensions for demonstration, a consensus should be clear that the $Q_{jt}$ will be a curve with small stiffness, similar illustrations appear in Qtran++\cite{son2020qtran++}. 



\subsection{Irrationality of OW-QMIX}
Next, we will use an OMG to experimentally demonstrate that OW-QMIX is an overly ideal algorithm.
We provide a payoff matrix with slightly increased difficulty, whose suboptimal values have a greater impact than the matrix used in QTRAN, QPLEX\cite{wang2020qplex}. But this is still an ordinary matrix, which satisfies a symmetric distribution, implying that the agents are isomorphic.
\begin{table}[H]
    \label{tab:Payoff of OMG}
    \setlength{\tabcolsep}{5mm}
    \caption{Payoff of OMG}
    \centering
    \begin{tabular}{|c|c|c|c|}
        \hline
        \diagbox{$\cal A$}{$\cal B$} & \textbf{A}&B&C\\
        \hline
        \textbf{A}&\textbf{8}&-12&-12\\
        \hline
        B&-12 & 6&6\\
        \hline
        C&-12&6&6\\
        \hline
    \end{tabular}
\end{table}

\begin{table}[!t]
\caption{Result of Weighted QMIX}
    \label{tabowqmixomg}
    \centering
    \begin{subfloat}{\setlength{\tabcolsep}{5mm}
        \begin{tabular}{|c|c|c|c|}
        \multicolumn{4}{c}{$\alpha=0.001$}\\
        \hline
        \diagbox{$\cal A$}{$\cal B$} & \textbf{0.267}&-0.022&0.001\\
        \hline
        \textbf{3.078}&\textbf{8.263}&7.057&7.123\\
        \hline
        -0.007&7.190 & 6.038&6.102\\
        \hline
        -0.005&7.237&6.083&6.147\\
        \hline
    \end{tabular}}
    \end{subfloat}\\
        \begin{subfloat}{\setlength{\tabcolsep}{5mm}
        \begin{tabular}{|c|c|c|c|}
        \multicolumn{4}{c}{$\alpha=0.01$}\\
        \hline
        \diagbox{$\cal A$}{$\cal B$} & \textbf{0.302}&0.003&0.000\\
        \hline
        \textbf{0.193}&\textbf{7.218}&6.278&6.183\\
        \hline
        0.008&6.768&5.833&5.739\\
        \hline
        0.007&6.743&5.808&5.715\\
        \hline
    \end{tabular}}
    \end{subfloat}\\
        \begin{subfloat}{\setlength{\tabcolsep}{5mm}
        \begin{tabular}{|c|c|c|c|}
        \multicolumn{4}{c}{$\alpha=0.1$}\\
        \hline
        \diagbox{$\cal A$}{$\cal B$} & -0.079&\textbf{0.144}&0.057\\
        \hline
        0.012&4.621&4.959&4.825\\
        \hline
        \textbf{0.032}&4.639&\textbf{4.979}&4.844\\
        \hline
        0.025&4.632&4.972&4.837\\
        \hline
    \end{tabular}}
    \end{subfloat}\\
        \begin{subfloat}{\setlength{\tabcolsep}{5mm}
        \begin{tabular}{|c|c|c|c|}
        \multicolumn{4}{c}{$\alpha=0.5$}\\
        \hline
        \diagbox{$\cal A$}{$\cal B$} &-4.174&0.383&\textbf{0.383}\\
        \hline
        -3.935&-7.652&-7.507&-7.507\\
        \hline
        0.376&-7.627& 5.715&5.716\\
        \hline
        \textbf{0.377}&-7.627&5.726&\textbf{5.727}\\
        \hline
    \end{tabular}}
    \end{subfloat}
\end{table}

Table. \ref{tabowqmixomg} shows the individual action values and joint action values of OW-QMIX under different weights. Boldface means the optimal individual actions and the optimal joint action. When $\alpha=0.5$, the optimal joint action combination is completely pinned by the surrounding minimal values. 
Even in the case of $\alpha=0.1$, OW-QMIX still cannot solve the value decomposition problem correctly.
$\alpha=0.01$ is a relatively suitable weight but also produces a little underestimation.
And when $\alpha=0.001$, all of the values are overestimated.
In summary, when a large weight is selected, its effect is no different from that of QMIX, and $Q_{jt} = y$ will never be obtained by the maximum values with most handcrafted weights. When a small enough weight is selected, $Q_{jt}$ corresponding to the larger value in $y$ may be out of control due to the small tension force (if you accept the analogy of the elastic bands): ${Q_{jt}} > y$ for all $\bm{a}$, and eventually lead to a catastrophic overestimation\cite{DDQN} of the whole Markov chains.

Conclusions can be drawn from the above experiment that Weighted QMIX does not change the fact that all gradients are linearly and positively related to the corresponding TD errors, and thus are still highly susceptible to the influence of minimum values. 
In practice, the distribution of $y$ is dynamic and unknown, making it difficult for the existing schemes to choose a suitable weight. Theoretical analysis of the impracticability can be found in Appendix. 

\section{MCVD}
In this section, we propose a novel value function decomposition method called MCVD, where the advantage of correntropy in filtering out outliers is transferred to the MARL domain. 
The key point is to break the linear positive correlation between the gradient and the corresponding TD error, and eliminate the negative influence generated by the minimal values on the maximum value fitting.

\subsection{Rethinking The Value Decomposition Problem}
Researchers are trying to solve the value decomposition problem, especially the non-monotonic value decomposition problem, through various means, such as complex network structures, inequalities, and segmented discussion, but few studies have analyzed its nature qualitatively.

The proposal of IGM makes the solution to the value decomposition problem appear to solve a matching relationship. However, with the existing technology, to prevent the divergence of the individual action value, we have to establish a partial equation relationship between the combination of individual action values and the joint action values to accomplish the matching task.
Through the analysis of VDN and QMIX, we find that the monotonic value decomposition problem can essentially be regarded as a regression problem, that is, the combinations of individual actions should be equivalent to the joint action values: $Q_{jt}=y$. The analysis of Weighted QMIX shows that the non-monotonic value decomposition problem can be further classified as a robust regression problem\cite{chen2017robust} due to the requirement to avoid the interference of the minimum joint action values.

Compared with the traditional robust regression problem, this problem also contains the following two characteristics:\\
1. In the joint action values, minimal ones are regarded as outliers, and maximal ones are regarded as target values, thus the regression problem exhibits the property of one-edged fitting.\\
2. According to the analysis in sec. \ref{im wqmix}, it can be known that $Q_{jt}$ is limited by summation or monotonicity constraints, thus it is a model with a serious lack of fitting ability. The regression problem also exhibits the property of irreconcilable underfitting.

Given the above properties, we characterize the non-monotonic value decomposition problem as an Underfitting One-edged Robust Regression problem\footnote{This characterization still holds for using summation to solve monotonicity problems.}. 
Under this definition, many robust regression methods can be applied to the value decomposition problem, which broadens the horizon of research on this problem.

According to our expertise, we make the first attempt to give a solution to the value decomposition problem from the perspective of information-theoretic learning\cite{principe2010entropy}.
The basic idea is to utilize the information theory descriptors of entropy as a cost function for the design of value decomposition algorithms. 
In the next subsection, we will introduce correntropy as a cost function into the field of value decomposition and hope to play a role in attracting more ideas to this interdisciplinary research.

\subsection{Maximum Correntropy Criterion}
First, we briefly introduce the basic concepts and effects of MCC.

Correntropy is a generalized similarity measure between two random variables. Given two random variables $X,Y \in \mathbb{R}$ with joint distribution function ${F_{XY}}(x,y)$, correntropy is defined by
\begin{equation}
    V(X,Y) = \mathbb{E}[\kappa (x,y)] = \int {\kappa (x,y)d} {F_{XY}}(x,y),
    \label{Eq:k-correntropy}
\end{equation}
where $\kappa ( \cdot , \cdot )$ is a shift-invariant Mercer Kernel. The most commonly used kernel is the Gaussian Kernel, given by
\begin{equation}
    \kappa (x,y) = {G_\sigma }(x,y) = \exp ( - \frac{{{{(x - y)}^2}}}{{2{\sigma ^2}}})
\end{equation}
where $\sigma  > 0$ stands for the kernel bandwidth.

\begin{figure}[]
    \centering
    \includegraphics[width = 3in]{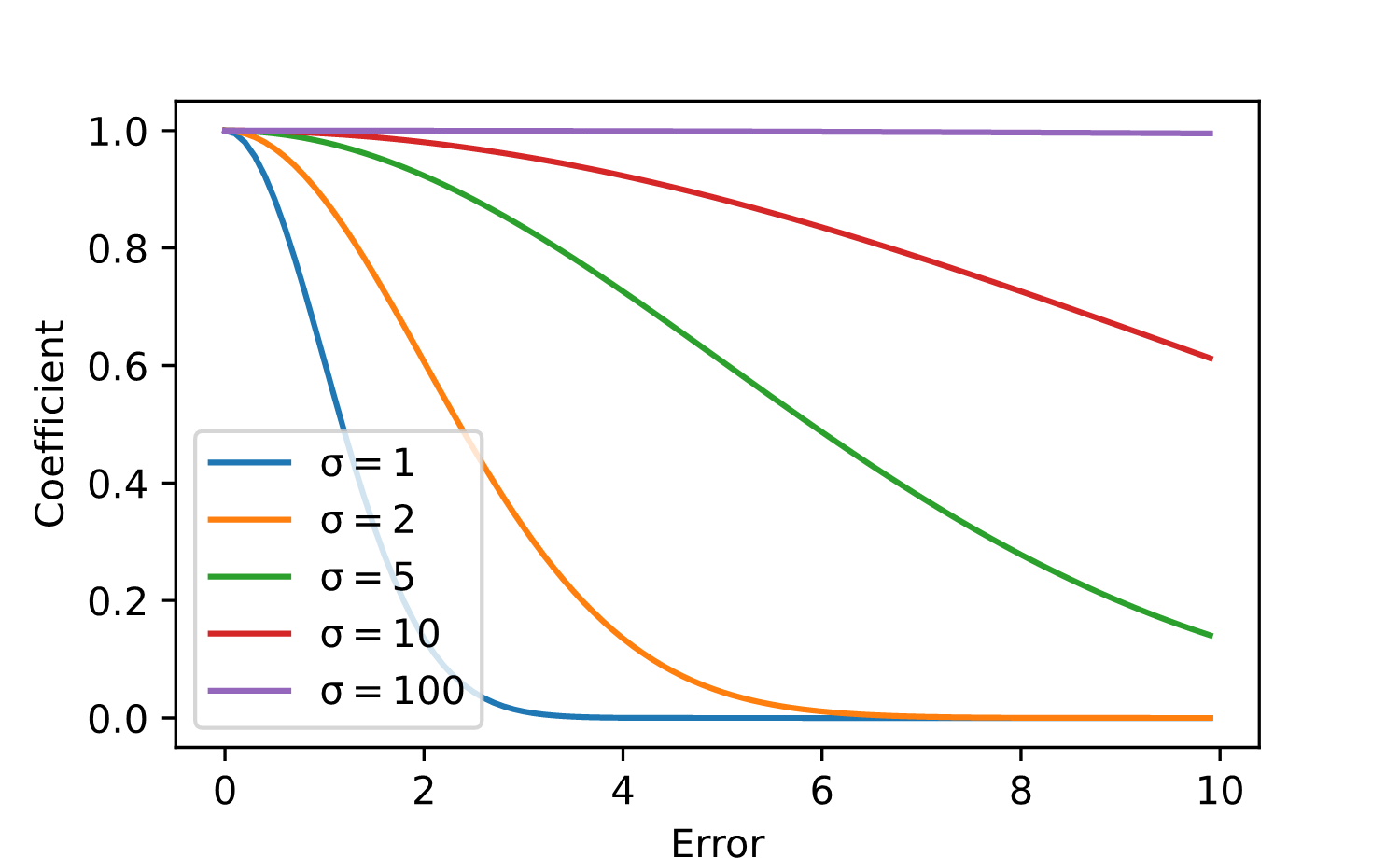}
    \caption{The relationship between coefficient, error and the kernel bandwidth. 
    A small kernel bandwidth combined with large errors would result in very small coefficients, consequently shrinking the gradients of large errors. When the kernel bandwidth is extremely large, or even, infinite, the coefficient would always be one no matter what the error is, then MCC loss degrades to MSE loss.}
    \label{fig:kernel bandwidth}
\end{figure}

Eq.\ref{Eq:k-correntropy} is the expression for continuous systems. In most realistic situations, however, only limited numbers of data are available, and the joint distribution ${F_{XY}}(x,y)$ is hard to estimate. So a practical operation is to estimate the correntropy using a sample mean estimator for discrete systems\footnote{Compared to some references, we multiply the coefficient term by $2\sigma ^2$ to achieve gradient normalization.}:
\begin{equation}
    \hat V(X,Y) = \frac{{2{\sigma ^2}}}{M}\sum\limits_{i = 1}^M {{G_\sigma }(x(i),y(i))} ,
\end{equation}
with ${\{{x(i),y(i)}\}^M_{i=1}}$ being $M$ samples drawn from $F_{XY}$. 
If we cal the gradient of $\hat V(\bm{e})$, where $\bm{e} = {e(1)} \times {e(2)} \times  \cdots  \times {e(n)}$, ${e(i)} = x(i) - y(i)$, assuming that $e(i),e(j)$ are independent identically distributed(iid), $\forall i,j \in b$:
\begin{equation}
    \frac{{\partial \hat V(\bm{e})}}{{\partial \bm{e}}} =  - \frac{2}{M}\sum\limits_{i = 1}^M {\exp ( - \frac{{e{{(i)}^2}}}{{2{\sigma ^2}}})e(i)} ,
\end{equation}
If we try to maximum the correntropy, where gradient ascent(denote as $\Delta$) is used instead of gradient descent(denote as $\nabla$). Take $\hat V(e)$ as the loss function ${{\cal L}}_{mcc}$, and the contrast of MCC loss ${{\cal L}}_{mcc}$ and traditional MSE loss ${{\cal L}}_{mse}$ is:
\begin{equation}
\label{eq mcc loss}
    \Delta {{{\cal L}}_{mcc}} = - \frac{{\partial {{{\cal L}}_{mcc}}}}{{\partial \bm{e}}} = \frac{2}{M}\sum\limits_{i = 1}^M {\exp ( - \frac{{e{{(i)}^2}}}{{2{\sigma ^2}}})e(i)} ,
\end{equation}
\begin{equation}
\label{eq mse loss}
    \nabla {{{\cal L}}_{mse}} = \frac{{\partial {{{\cal L}}_{mse}}}}{{\partial \bm{e}}} = \frac{2}{M}\sum\limits_{i = 1}^M {e(i)} ,
\end{equation}
Compared with MSE loss, the gradient of MCC loss add an exponential coefficient to each error, i.e. the Gaussian Kernel function. The relationship between coefficient, error, and the kernel bandwidth is shown in Fig. \ref{fig:kernel bandwidth}.

The error and coefficient show a negative correlation under MCC loss, thus the interference of outliers in the system can be greatly weakened.

\subsection{Implementation of MCVD}
To eliminate the irrationality of the weight design of OW-QMIX, we focus on breaking the linear positive correlation between gradients and corresponding TD errors.
Our approach is to treat the minimal values as outliers that should be filtered out, establishing a rationale for migrating MCC to the non-monotonic value decomposition problem.

In order to maximize the TD errors of the minimum values, $Q_{jt}$ needs to be raised first, so that the MCC loss can function correctly, this process is the same as Weighted QMIX. Then followed by traction and correction for $Q_{jt}$ using MCC loss.
With a solid foundation for Weighted QMIX and MCC, the loss function of MCVD will be very easy for understanding:
\begin{equation}
\begin{array}{l}
{{{\cal L}}_{td}} = \left\{ {\begin{array}{*{20}{l}}
{{{{\cal L}}_{mse}}({Q_{jt}}(s,\bm{a}),y),}\\
{ - {{{\cal L}}_{mcc}}({Q_{jt}}(s,\bm{a}),y),}
\end{array}} \right.\begin{array}{*{20}{l}}
{if\quad{Q_{jt}}(s,\bm{a}) \le y}\\
{others}
\end{array},\\
{{{\cal L}}_{mcc}}({Q_{jt}}(s,\bm{a}),y) = {\mathbb{E}_{s,\bm{a} \sim b}} {[{2\sigma ^2}\exp ( - \frac{{{{({Q_{jt}}(s,\bm{a}) - y)}^2}}}{{2{\sigma ^2}}})]} ,\\
{{{\cal L}}_{mse}}({Q_{jt}}(s,\bm{a}),y) = {\mathbb{E}_{s,\bm{a} \sim b}} {[{{({Q_{jt}}(s,\bm{a}) - y)}^2}]} ,
\end{array}
\end{equation}
In order to further simplify and make it look more similar to the original expression of Weighted QMIX, we perform partial gradient pruning from the code level. 
By comparing Eq. \ref{eq mcc loss}, \ref{eq mse loss} and detaching the gradient of the coefficient term, the following expression can be easily obtained:
\begin{equation}
\begin{array}{l}
{w_{td}} = \exp ( - \frac{{({Q_{jt}}(s,\bm{a}) - y)_{clip(\min  = 0)}^2}}{{2{\sigma ^2}}}),\quad detached\\
{{{\cal L}}_{td}} = {\mathbb{E}_{s,\bm{a} \sim b}} {[{w_{td}}{{({Q_{jt}}(s,\bm{a}) - y)}^2}]} ,
\end{array}
\end{equation}

After the above deformation, to our best knowledge, MCVD becomes the simplest one of the recent value decomposition algorithms.
We use Fig. 3 to further illustrate how MCVD works and how it differs from OW-QMIX.
\begin{figure}[]
    \centering
    \includegraphics[width = 3in]{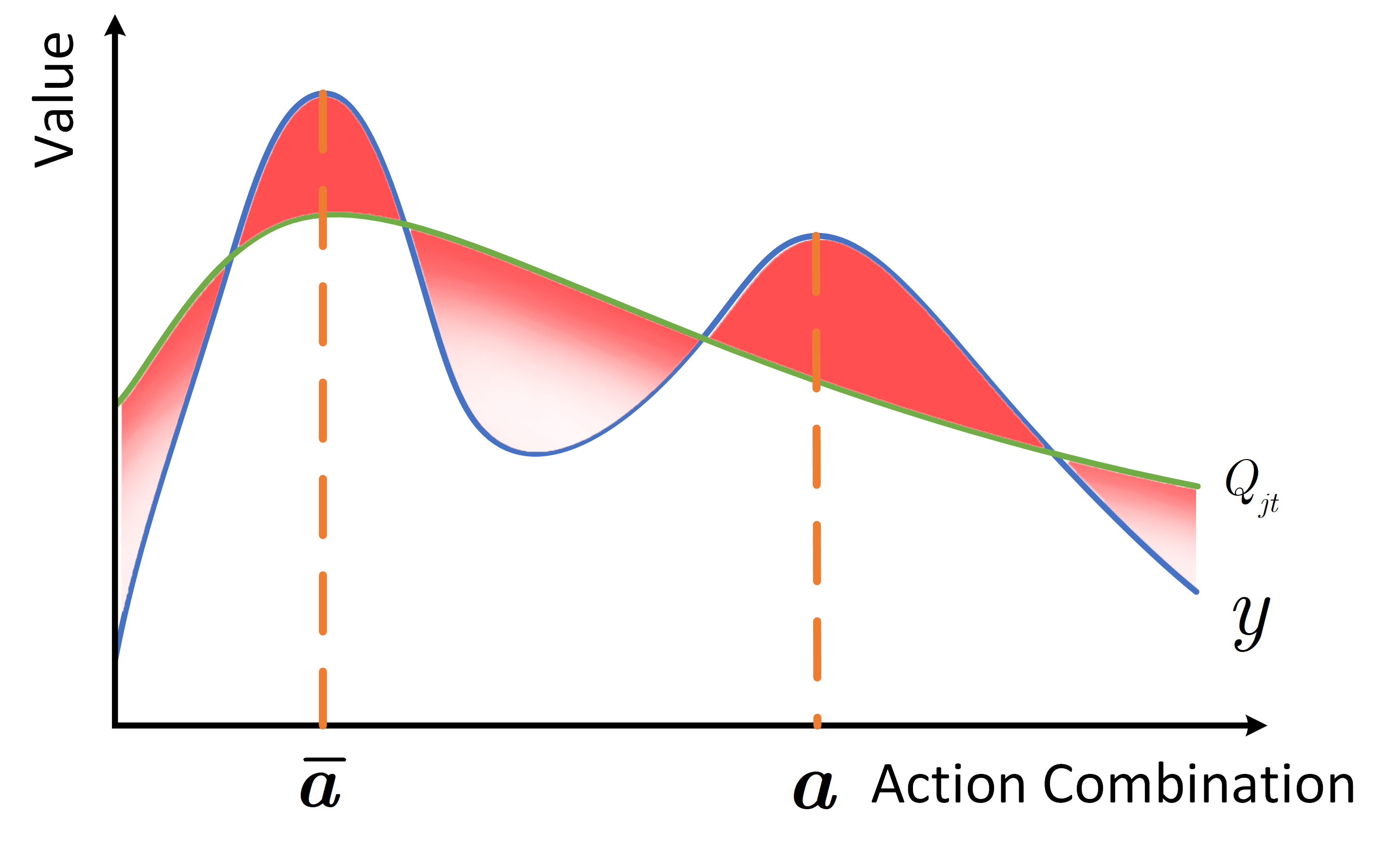}
    \caption{Illustration of MCVD, where the saturation represents the value of the weight, the lower the saturation, the smaller the weight.}
    \label{fig:ill-mcvd}
\end{figure}

MCVD breaks the linear positive correlation between gradients and TD errors, and realizes a smooth decrease of the weight with the increase of the TD error, thereby eliminating the negative impact of the minimum values to a certain extent. Theoretical analysis of the rationality can be found in Appendix.

\subsection{Details of MCVD}
\label{subsec details mcvd}
In this subsection, we introduce two design details of MCVD. 
The first point is that we use the sum of individual action values instead of the complex hyper-mixing network, which is why our proposed algorithm is called MCVD instead of MC-QMIX.

The second point is that we use a joint action-value approximation network for joint-action value approximation, the pros and cons of the trade-offs for selecting various networks will be analyzed.

\subsubsection{Selection of Value Decomposition Network}
After the advent of QMIX, many derivative algorithms\cite{SMIX,UNMAS,son2020qtran++} are based on the hyper-network to establish the connection between joint action-values and individual action values.
However, in MCVD, the joint action-values is obtained by summing all individual action values, just like VDN does.
We argue that there are pros and cons to using summation and positively weighted summation for non-monotonic value decomposition problems.
A positively weighted summation can certainly extend the expressiveness of the joint action-value function, but it is not necessary for the problem we are trying to solve.
The goal that we and many researchers hope to achieve, is that ${Q_{jt}} \ge y$, and according to the IGM condition, we only need to focus on the global and local optimal consistency, and there is no need to care about the relationship between non-optimal joint action-values and non-optimal individual action values.
Traditional methods need to achieve an equal relationship between ${Q_{jt}}$ and $y$, so a richer expression is necessary for a better result. 
But for inequalities, hyper-network becomes a burden for learning.
Ablation experiments are conducted and demonstrated that the use of hyper-network slows the learning, and the performance has no sign of improvement. Detailed results can be found in Appendix.

\subsubsection{Discussion of joint action-value Approximation Network}
We use the same joint action-value approximation network ${{\hat Q}_{jt}}(s,\bm{a})$ as in QTRAN for the iteration of Bellman equation. 
\begin{equation}
\begin{array}{l}

{{\cal L}_{jt}} = {\mathbb{E}_{s,\bm{a} \sim b}} {[{{({{\hat Q}_{jt}}(s,\bm{a}) - y)}^2}]},\\
y = r + \gamma (1-t){{\hat Q'}_{jt}}(s,\bm{\bar u}){|_{\bm{\bar u'} = \{ \mathop {\arg \max }\limits_{u_i} Q_i'(o', \cdot )\} }},
\end{array}
\end{equation}
Existing algorithms with joint action-value approximation networks, such as QTRAN and CW-QMIX, have not achieved excellent results. 
We suppose that the reason for the unsatisfactory performance is that these algorithms do not pick out the real optimal combination of individual action values, rather than the use of joint action-value network.
We are in favor of distinguishing the joint action-value approximation function for Bellman Equation iteration from the value decomposition network for optimal action learning, to reduce the coupling effect and avoid gradient interaction.

It is also interpretable that the value decomposition alone can achieve better results.
As can be seen from Tab. 2 that most learning results of OW-QMIX, ie. $Q_{jt}$ is a distribution with small variance, as we mentioned before, the combination of individual action values is a curve with small stiffness, so when the optimal action combination is transferred, iteration with the output of value decomposition network for Bellman Equation will make the learning smoother. 
However, the result of the joint action-value $y$ is like a mountain map with ravines, it is a distribution with large variances. When the optimal individual action value is transferred, the corresponding joint action value may change severely, and the entire Markov chain will have large fluctuations, which significantly increases the instability of learning.

To sum up, the learning result of the value decomposition network at each state is a relatively static distribution with small variance.
And the learning result of the joint action-value approximation network at each state is a relatively dynamic distribution with large variance.

However, using the results of the value decomposition network for the iteration of the Bellman Equation also has many disadvantages.
The first is that the intersection of gradients will increase the difficulty of learning, and the learning of individual action value networks will encumber the learning of the value decomposition network, thus the approximation speed and effect are worse than the joint action-value approximation network.
Secondly, the change of each individual action value will affect the result of the value decomposition network. Considering that in a multi-agent system, once sampling can only cover a small action combination space, it may take a long time for a wrong change to be corrected.
But since the joint action-value network uses the index of the action instead of the value of action, it suffers fewer effects from the above two points.
Finally and most importantly, even if the optimal combination of individual action values can be chosen correctly, the result of the value decomposition network is usually overestimated or underestimated, see Tab. \ref{tabowqmixomg}, which is severely detrimental to the iteration of the Bellman Equation\cite{DDQN,fujimoto2018addressing}. 
Combined with the disadvantage of the above two points, it is difficult for the agents to make the right decisions in a long Markov chain.

After the above analysis, we believe that using the joint action-value network for Bellman Equation iteration generally has more advantages than disadvantages.
It modularizes the entire system and uses its own output results to perform primitive reinforcement learning similar to a single agent, allowing the value decomposition network to focus on the gradient transmission for individual action-value networks. 
Even if the result of the value decomposition network is overestimated or underestimated, as long as the selected optimal action combination is correct, it will not have a negative impact on the Markov chain.
Ablation experiments are conducted and proved the improvement of using a joint action-value network, detailed results can be found in Appendix.

\subsection{Overall Framework of MCVD}
So far, we have presented our propositions sporadically, but have not yet formed a systematic framework for practice. In this chapter, we will give the architecture diagram and pseudo-code of MCVD macroscopically.

\begin{figure*}[!t]
    \centering
    \includegraphics[width = 0.8\textwidth]{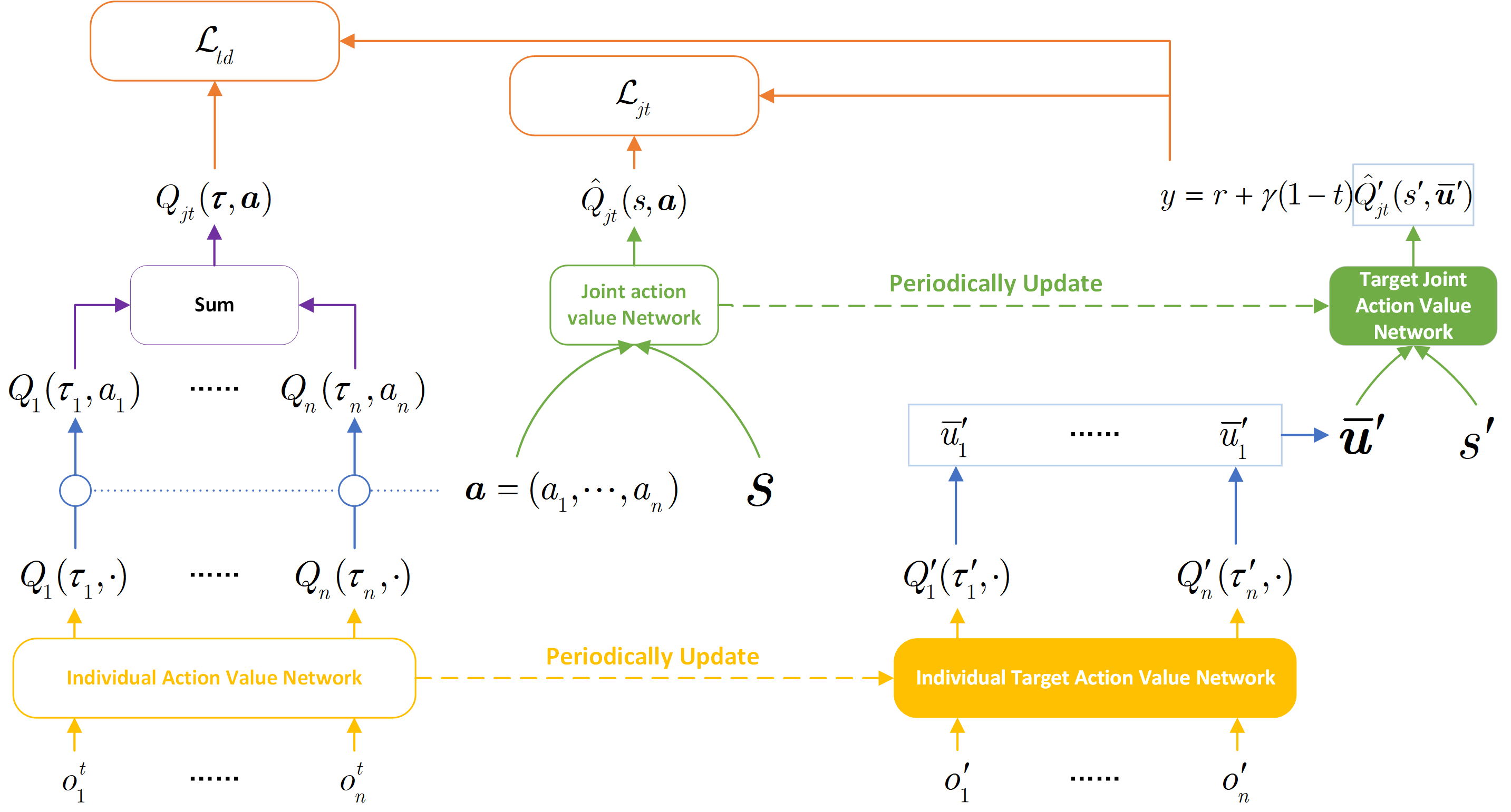}
    \caption{The architecture of MCVD.}
    \label{fig:mcvd-overview}
\end{figure*}

The architecture of MCVD is shown in Fig. 4, and its structure is relatively simple, containing two online networks\footnote{It is usually represented as the eval network when writing codes. The online network is born from Double DQN.}, which are the individual action value network and the joint action-value network, and the corresponding two target networks.
Two loss functions are utilized to update the above two online networks respectively.

The pseudo-code of MCVD is shown in Algorithm 1.

\begin{algorithm}
\caption{MCVD}
\begin{algorithmic}
\STATE\textbf{Initialize} networks, set the kernel bandwidth $\sigma $,

\STATE \textbf{For} episode $0$ to max-episodes $T$,
\STATE \hspace{0.5cm}obtain ${s, \bm{o}}$ from the env,
\STATE \hspace{0.5cm}\textbf{For} step 0 to max-episode-step $T_s$,
\STATE \hspace{1cm}\textbf{For} $i=1$ to $N$,
\STATE \hspace{1.5cm}${a_i} = \mathop {\arg \max }\limits_a ({Q_i}({o_i}, \cdot |),$
\STATE \hspace{1cm}${s} \times {\bm{a}} \mapsto {s'}$, get ${{\bm{o'}},{r}}$
\STATE \hspace{1cm}Store experience tuple $({s},\bm{o},\bm{a},{r},{s'},\bm{o'})$ to the replay 
\STATE \hspace{1cm}buffer ${{\cal D}}$,
\STATE \hspace{1cm}update ${s} \leftarrow {s'}$.
\STATE \hspace{0.5cm}\textbf{If} terminated and episodes mod train-frequency,
\STATE \hspace{1cm}sample a mini-batch $(s,\bm{o},\bm{a},r,s',\bm{o'})$ from the 
\STATE \hspace{1cm}replay buffer ${{\cal D}}$,
\STATE \hspace{1cm}Calculate:
\STATE \hspace{1cm}\textbf{For} $i=1$ to $N$,
\STATE \hspace{1.5cm}${{\bar u'}_i} = \mathop {\arg \max }\limits_u ({{Q'}_i}({{o'}_i}, \cdot  )),$
\STATE \hspace{1cm}$y = r + \gamma (1 - t){{\hat Q'}_{jt}}(s',\bm{\bar u'}),$\\
\STATE \hspace{1cm}${Q_{jt}}(\bm{o},\bm{a}) = \sum\limits_{i = 1}^{{N}} {{Q_i}({o_i},{a_i})} ,$\\
\STATE \hspace{1cm}${w_{td}} = \exp (\frac{{({Q_{jt}}(\bm{o},\bm{a}) - y)_{clip(\min  = 0)}^2}}{{2{\sigma ^2}}}),$\\
\STATE \hspace{1cm}Update the individual action value network:
\STATE \hspace{1cm}${{{\cal L}}_{td}} = {\mathbb{E}_{s,\bm{a} \sim b}} {[{w_{td}}{{({Q_{jt}}(\bm{o},\bm{a}) - y)}^2}]} ,$.
\STATE \hspace{1cm}Update the joint action-value network:
\STATE \hspace{1cm}${{{\cal L}}_{jt}} = {\mathbb{E}_{s,\bm{a} \sim b}} {[{{({\hat{Q}_{jt}}(\bm{s},\bm{a}) - y)}^2}]} ,$
\STATE \hspace{1cm}\textbf{If} episodes mod delay-frequency,
\STATE \hspace{1.5cm}Update target networks.
\STATE \textbf{end}
\end{algorithmic}
\end{algorithm}

\section{Experiments}
In this section, we conduct our experiments in three environments:\\
1. One-Step Matrix Game, which is a preliminary test of the non-monotonic value decomposition algorithm.\\
2. Cooperative-Navigation.
We modified the Cooperative-Navigation scenario of multi-agent particle environments (MPE)\cite{lowe2017multi} into a form similar to the SMAC interface.
The reward function of this scenario follows a non-monotonic distribution.
It is an effective experimental benchmark for verifying algorithms with non-monotonic decomposition ability.\\
3. StarCraft II micromanagement tasks. This environment is currently ruled by QMIX and Weighted QMIX\footnote{We do not take into account algorithms that use special techniques, such as RODE\cite{wang2020rode}}, whose reward functions do not follow non-monotonic distributions with large variances.

In contrast to Weighted QMIX, which uses different weights for each environment, in SMAC ($w=0.5$), and Predator-Prey ($w=0.1$), we choose a uniform kernel bandwidth ($\sigma=1$) for all scenarios to demonstrate the usability of MCVD.
Other hyperparameters are detailed in Appendix.

\subsection{OMG}
We conduct a preliminary experiment using the payoff matrix shown in Tab. 1 and select multiple kernel bandwidths to verify that the selection of kernel bandwidth in MCVD is more inclusive than the selection of weight in Weighted QMIX.

\begin{table}[]
\caption{Result of MCVD}
    \label{tab result mcvd OMG}
    \centering
    \begin{subfloat}{\setlength{\tabcolsep}{5mm}
        \begin{tabular}{|c|c|c|c|}
        \multicolumn{4}{c}{$\sigma=1$}\\
        \hline
        \diagbox{$\cal A$}{$\cal B$} & \textbf{4.470}&3.317&3.318\\
        \hline
        \textbf{3.483}&\textbf{7.935}&6.800&6.810\\
        \hline
        2.650&7.120 & 5.967&5.968\\
        \hline
        2.654&7.124&5.971&5.972\\
        \hline
    \end{tabular}}
    \end{subfloat}\\
        \begin{subfloat}{\setlength{\tabcolsep}{5mm}
        \begin{tabular}{|c|c|c|c|}
        \multicolumn{4}{c}{$\sigma=2$}\\
        \hline
        \diagbox{$\cal A$}{$\cal B$} & \textbf{4.506}&3.372&3.357\\
        \hline
        \textbf{3.425}&\textbf{7.931}&6.797&6.782\\
        \hline
        2.574&7.080&5.946&5.931\\
        \hline
        2.597&7.103&5.969&5.954\\
        \hline
    \end{tabular}}
    \end{subfloat}\\
        \begin{subfloat}{\setlength{\tabcolsep}{5mm}
        \begin{tabular}{|c|c|c|c|}
        \multicolumn{4}{c}{$\sigma=5$}\\
        \hline
        \diagbox{$\cal A$}{$\cal B$} & \textbf{4.317}&3.344&3.344\\
        \hline
        \textbf{3.647}&\textbf{7.964}&6.991&6.991\\
        \hline
        2.667&6.984&6.011&6.011\\
        \hline
        2.660&6.977&6.004&6.004\\
        \hline
    \end{tabular}}
    \end{subfloat}\\
        \begin{subfloat}{
        \setlength{\tabcolsep}{5mm}
        \begin{tabular}{|c|c|c|c|}
        \multicolumn{4}{c}{$\sigma=10$}\\
        \hline
        \diagbox{$\cal A$}{$\cal B$} & -0.466&0.259&\textbf{0.560}\\
        \hline
        -2.478&-2.944&-2.219&-1.918\\
        \hline
        2.057&1.591&2.316&2.617\\
        \hline
        \textbf{2.651}&2.185&2.910&\textbf{3.211}\\
        \hline
    \end{tabular}}
        \end{subfloat}
\end{table}

It can be seen from Tab. \ref{tab result mcvd OMG} that under most small kernel bandwidths, MCVD can perfectly solve the non-monotonic value decomposition problem with large variance distribution.
Moreover, its approximation effect on the optimal joint action-value is also better than that of Weighted QMIX, showing that both overestimation and underestimation are not obvious, and it also has a better approximation to the suboptimal values.
MCVD approaches the goal of $Q_{jt} \ge y$ as never before. By introducing MCC to the non-monotonic value decomposition field, the interference of the minimal values on learning is effectively eliminated.

Considering the large-scale non-monotonic value decomposition problem, $Q_{jt} = y$ is only satisfied for a very little number of values.
Theoretically, assuming that the action total of each agent is ${\left| A \right|}$, then only ${\left| A \right|}$ individual action combinations can perfectly satisfy that $Q_{jt} = y$ in the most extreme case, where the total of the joint action value is ${{\left| A \right|}^N}$.
Therefore, to a certain extent, a small kernel bandwidth in MCVD can adapt to environments with a large number of agents.
If we define a new indicator, Applicable Scope of the Kernel Bandwidth (ASKB), as the difference between the corresponding error and zero, where the corresponding error is the value when the coefficient of MCC loss with a certain kernel bandwidth satisfies $\exp ( - \frac{{e{{(i)}^2}}}{{2{\sigma ^2}}}) = \varepsilon ,$ and $\varepsilon$ is a very small positive constant: $\varepsilon>0$.
As can be seen from Fig. 2 that when the kernel bandwidth is small, the ASKB is small, too. Therefore, the kernel bandwidth cannot be infinitely small, and in an extreme case, $\sigma=0$, MCC fails for all errors.
Considering the above analysis, as well as the application of MCC in filtering, we finally choose $\sigma=1$ for subsequent experiments.

\subsection{Cooperative-Navigation}
Cooperative Navigation is a scene of the MPE, which is originally used as a  benchmark for MARL with continuous actions. 
We added state and executable actions to the original Cooperative Navigation, as well as a series of other interfaces so that it can be used by most MARL algorithms with discrete action space.
At the same time, in order to uniform the return curve when the number of agents is different, we average the reward function, that is, divide the originally designed reward by the corresponding number of agents, so that it does not increase or decrease linearly with the increase of the number of agents.
Under this improvement, regardless of the number of agents, the maximum value of the return curve should be close to zero, which is convenient for observation and comparison.

\begin{figure}[]
    \centering
    \subfloat[Six agents]{\includegraphics[width = 2.3in]{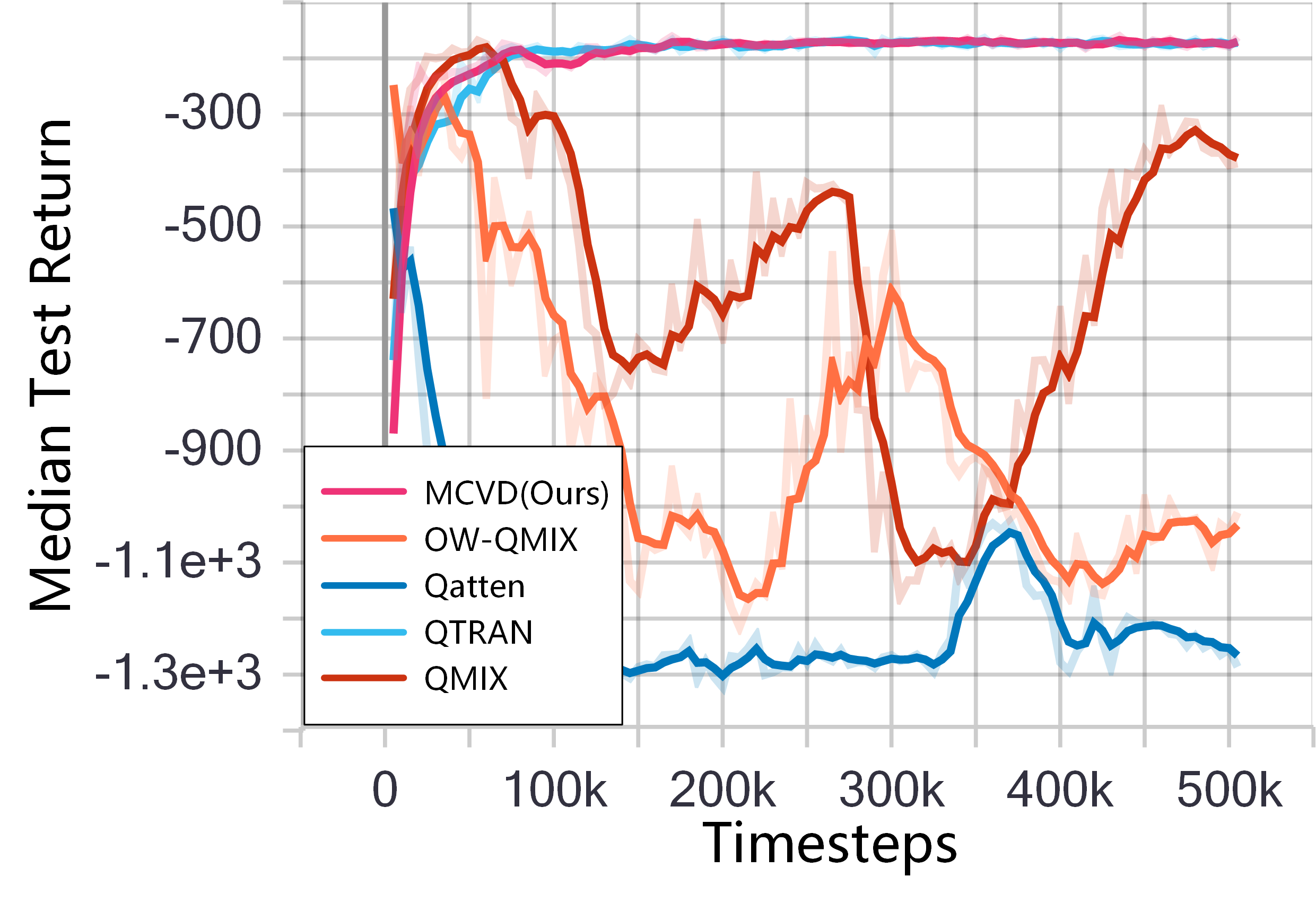}}\\
    \subfloat[Nine agents]{\includegraphics[width = 2.3in]{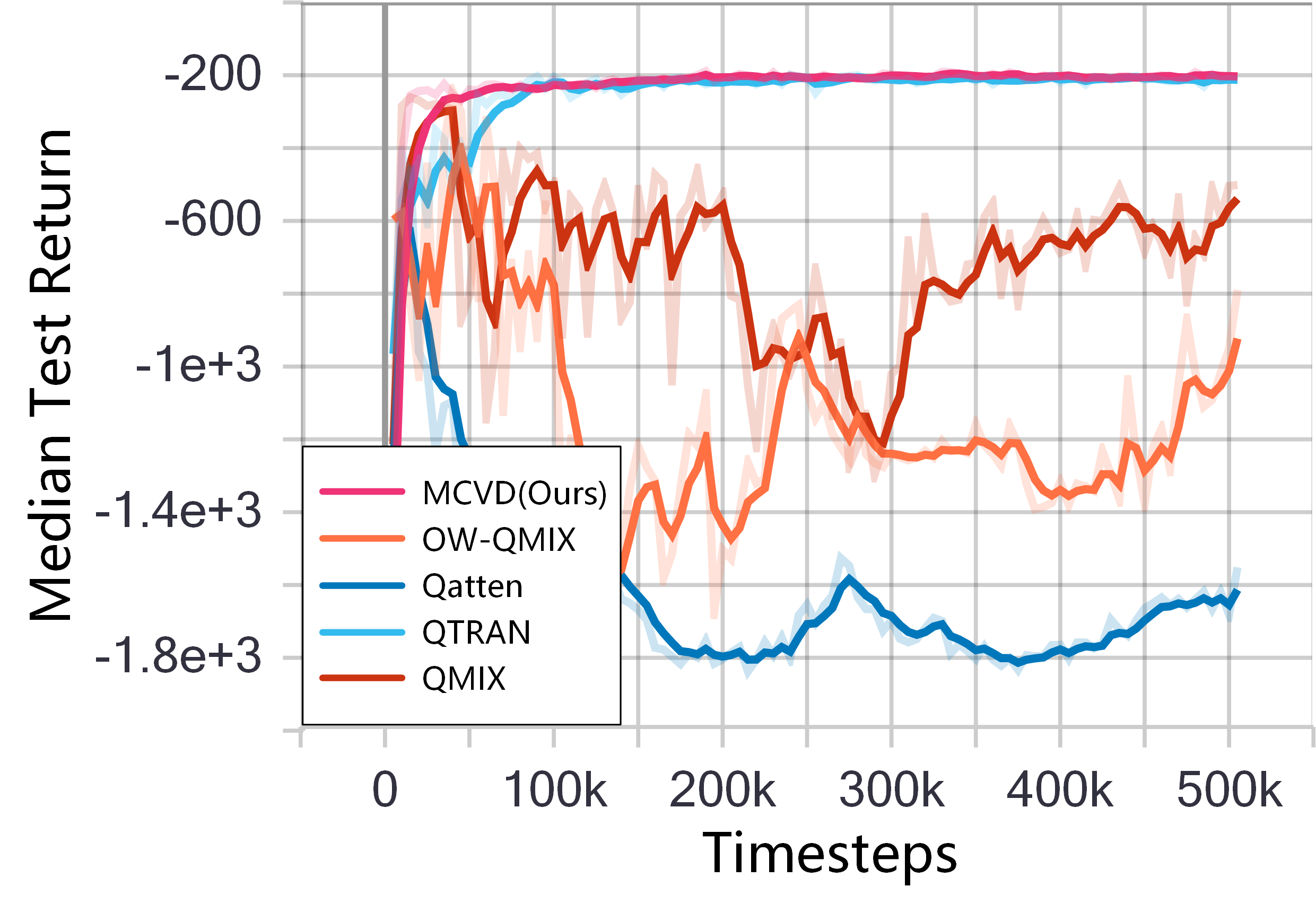}}
    \caption{Test return curves with different numbers of agents in Cooperative-Navigation scenario.}
    \label{fig:cn}
\end{figure}

We conducted experiments in the Cooperative-Navigation scenario with six and nine agents.
To get a sound performance of OW-QMIX, we set the weight $w=0.1$, which is consistent with the original setting in the Predator-Prey scenario.
From Fig. \ref{fig:cn}, we can see that only our proposed algorithm, MCVD, and QTRAN can converge in this scenario.
Meanwhile, at the beginning of training, MCVD learns faster than QTRAN.

This experiment confirms that our improved Cooperative-Navigation scenario is an effective non-monotonic value decomposition validation platform, and the proof of its non-monotonicity is given in the Appendix.
This experiment further confirms that MCVD is an effective non-monotonic value decomposition algorithm.
\subsection{SMAC}
We further conduct experiments in multiple SMAC scenarios to demonstrate the applicability of MCVD in broader environments.
\begin{figure*}[!t]
    \centering
    \subfloat[2s3z]{\includegraphics[width = 2.3in]{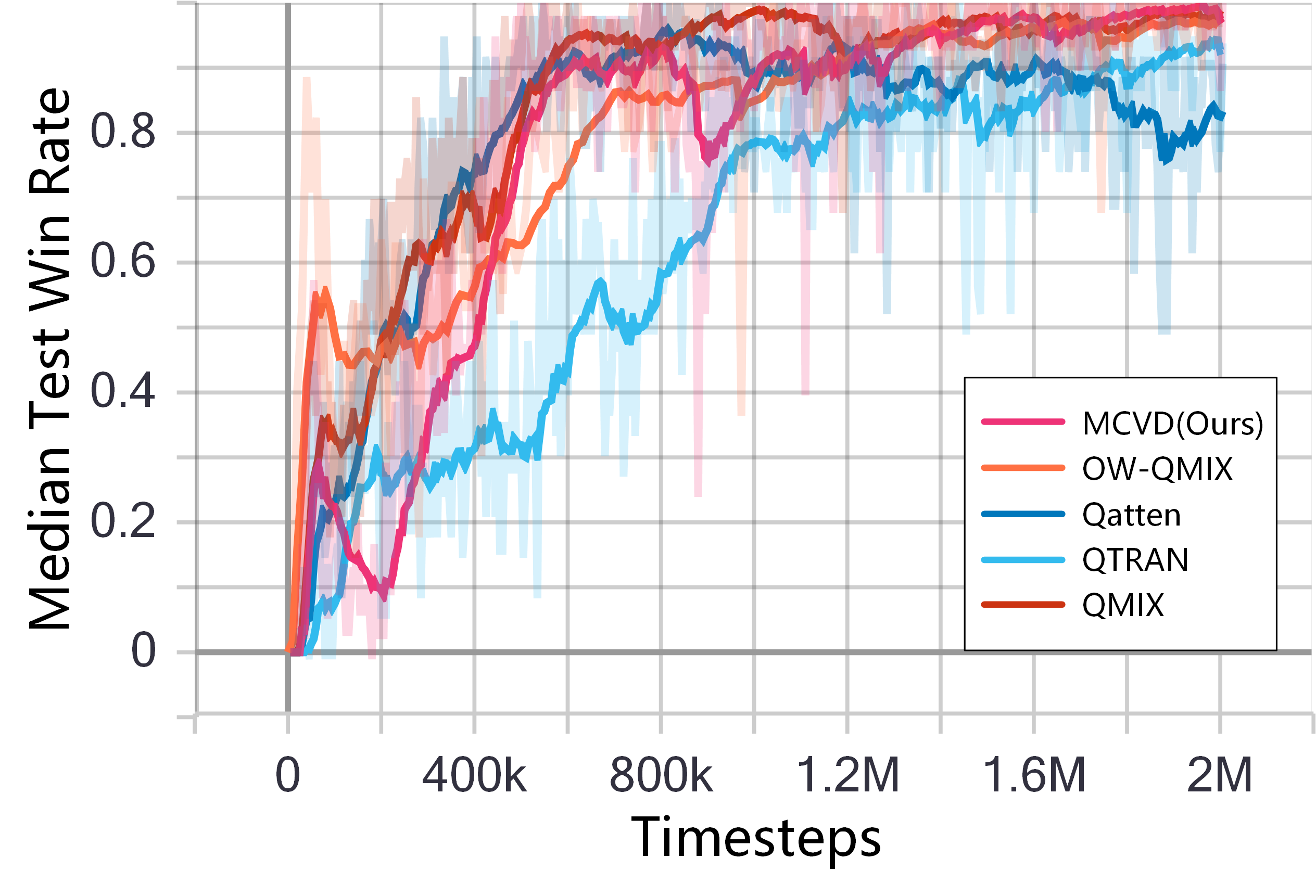}}
    \subfloat[3s5z]{\includegraphics[width = 2.3in]{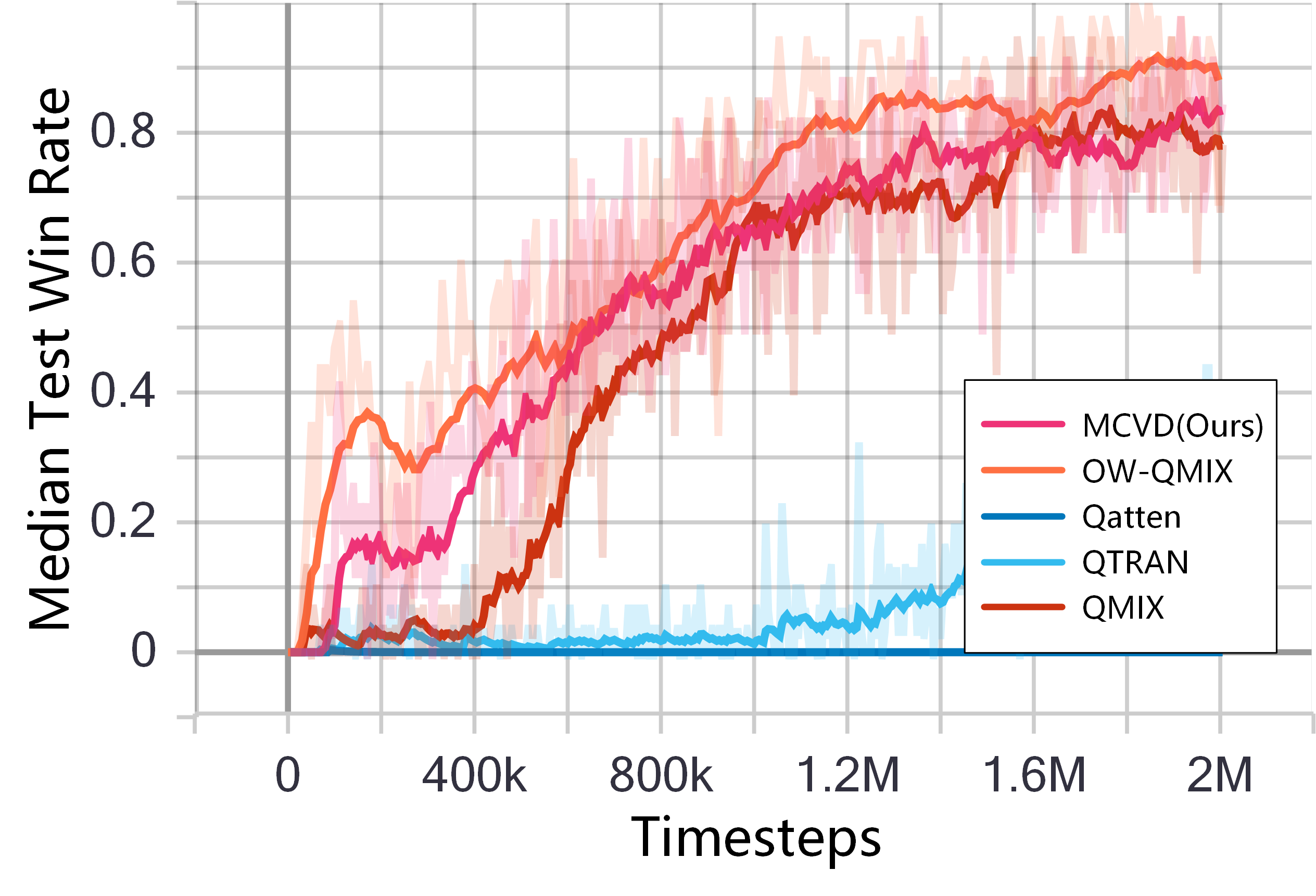}}
    \subfloat[1c3s5z]{\includegraphics[width = 2.3in]{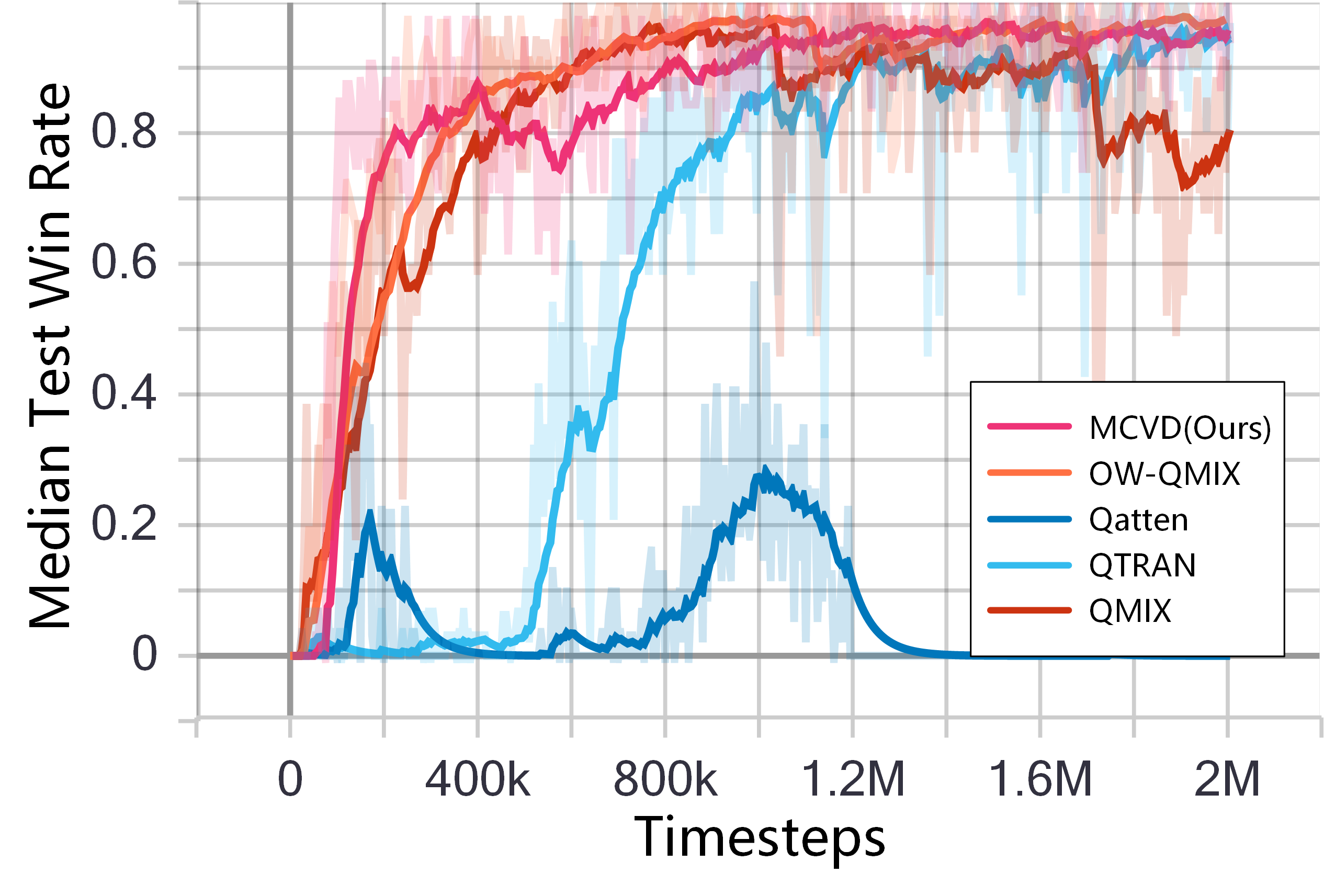}}\\
    \subfloat[2c\_vs\_64zg]{\includegraphics[width = 2.3in]{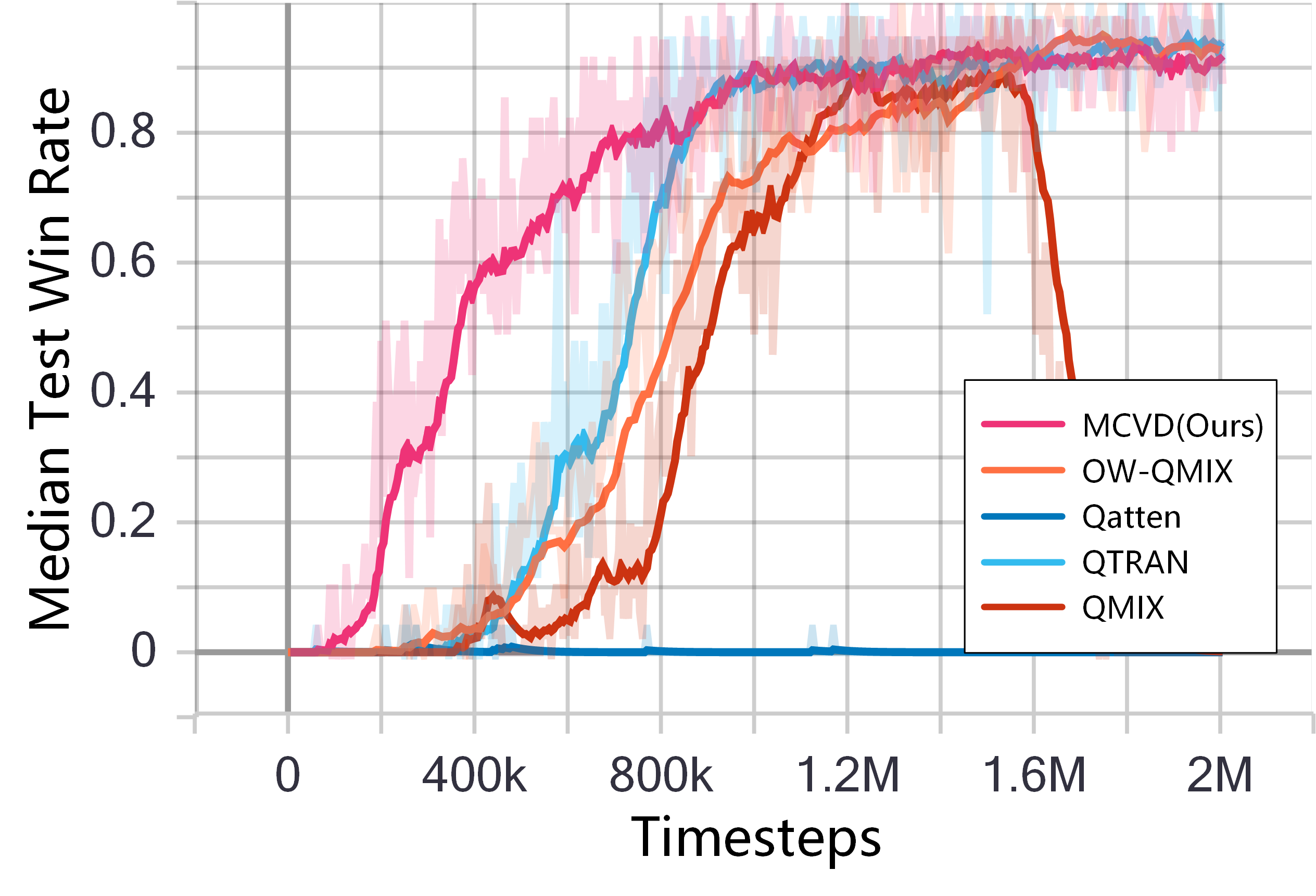}}
    \subfloat[3s\_vs\_5z]{\includegraphics[width = 2.3in]{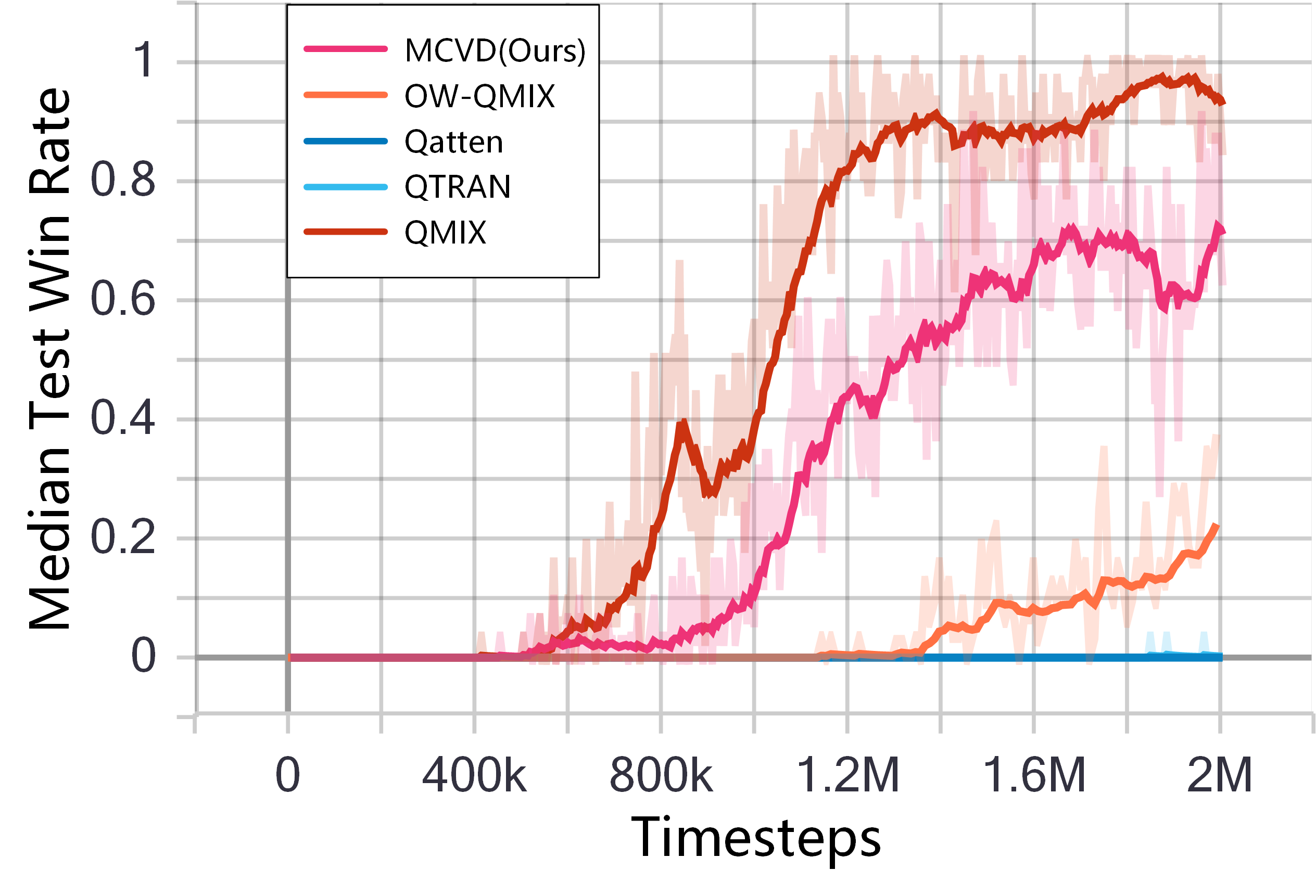}}
    \subfloat[5m\_vs\_6m]{\includegraphics[width = 2.3in]{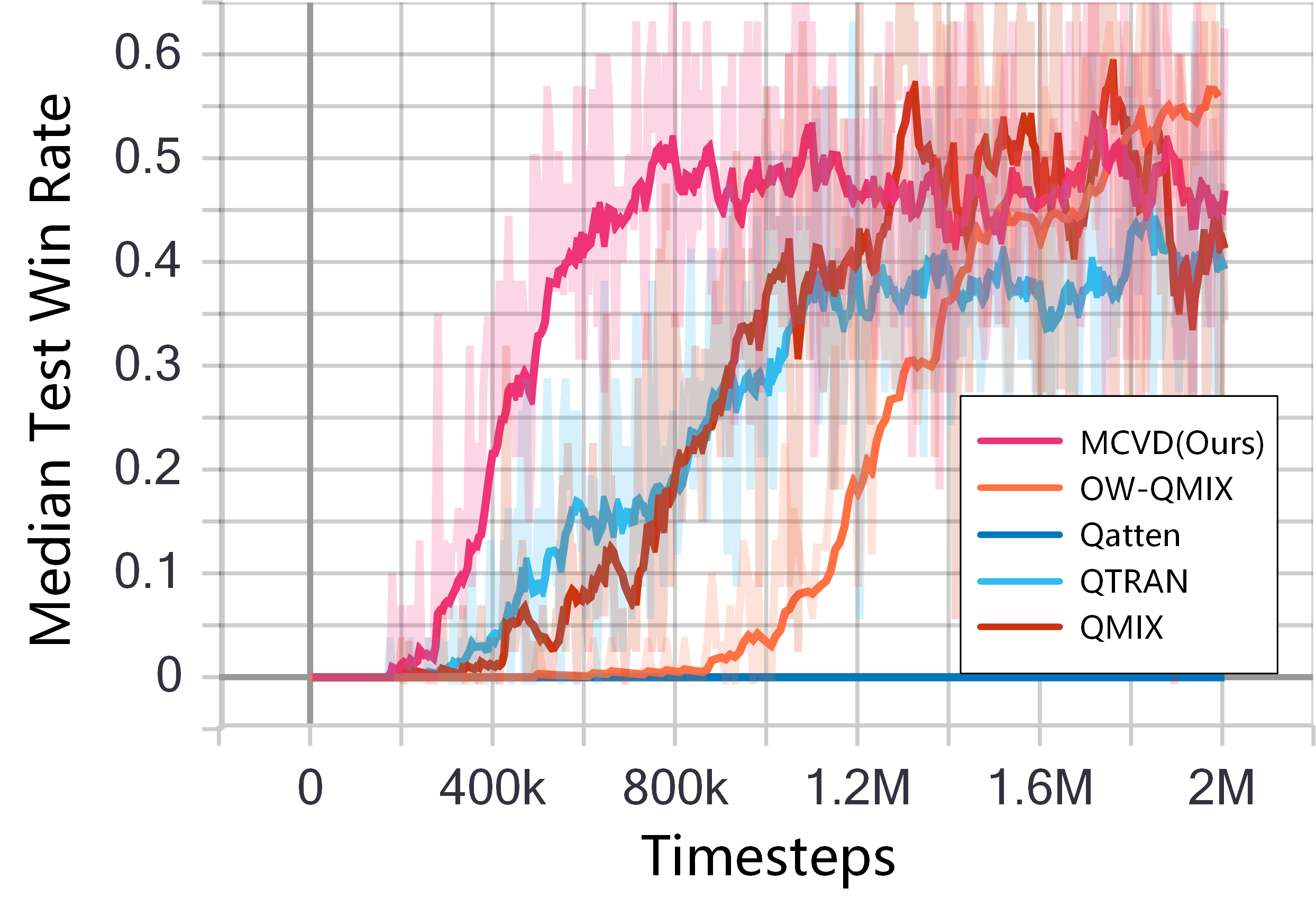}}\\
    \subfloat[8m]{\includegraphics[width = 2.3in]{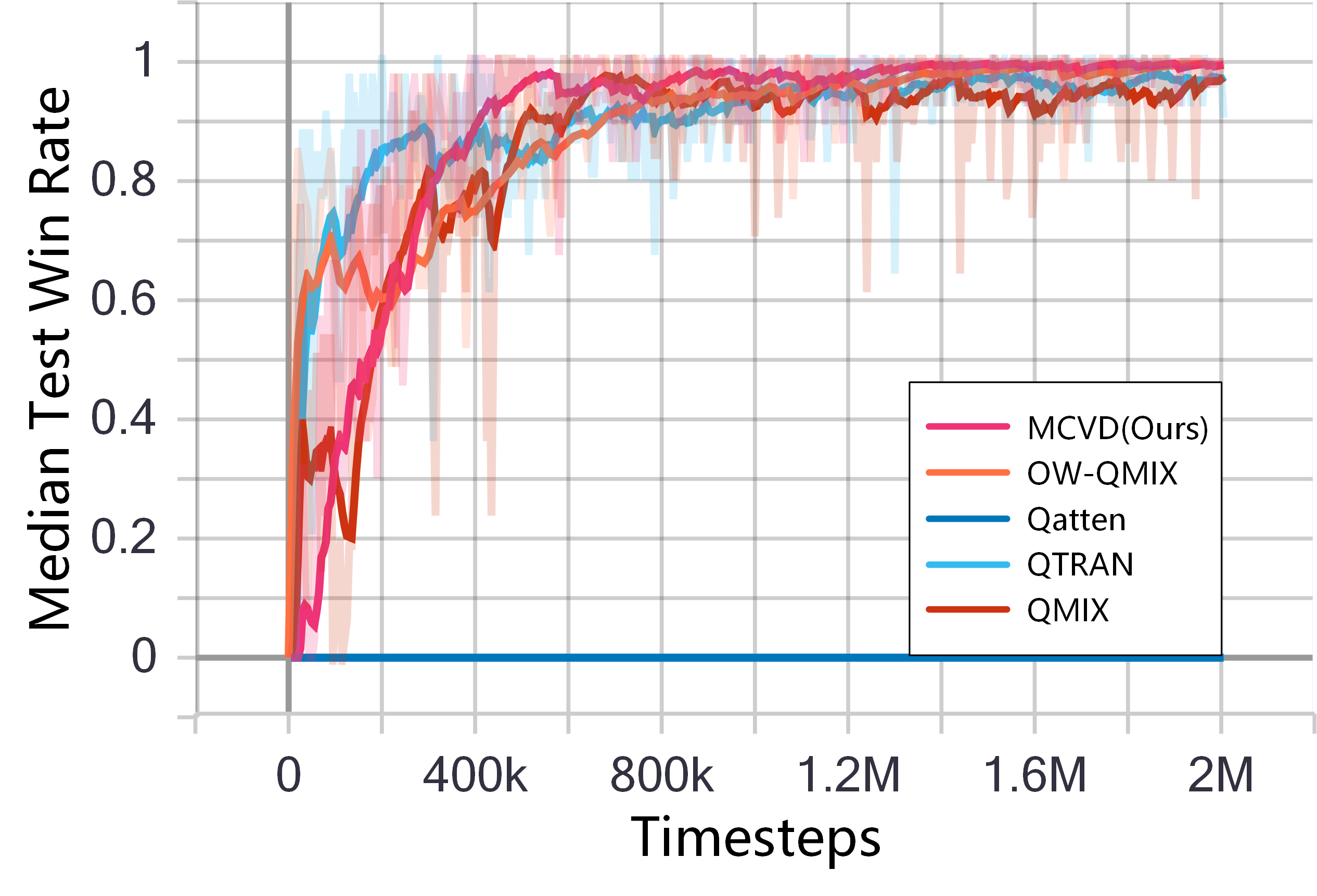}}
    \subfloat[10m\_vs\_11m]{\includegraphics[width = 2.3in]{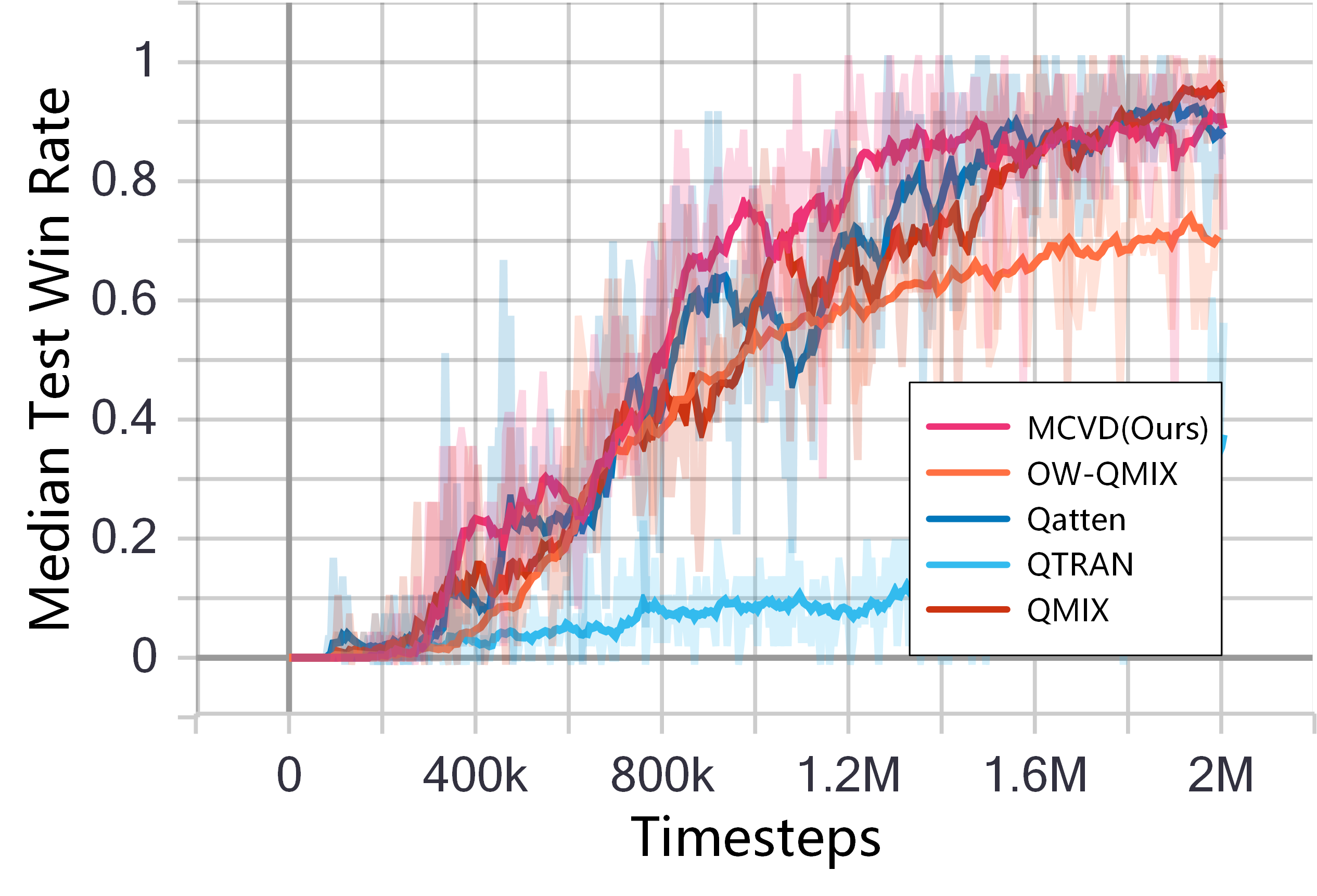}}
    \subfloat[MMM2]{\includegraphics[width = 2.3in]{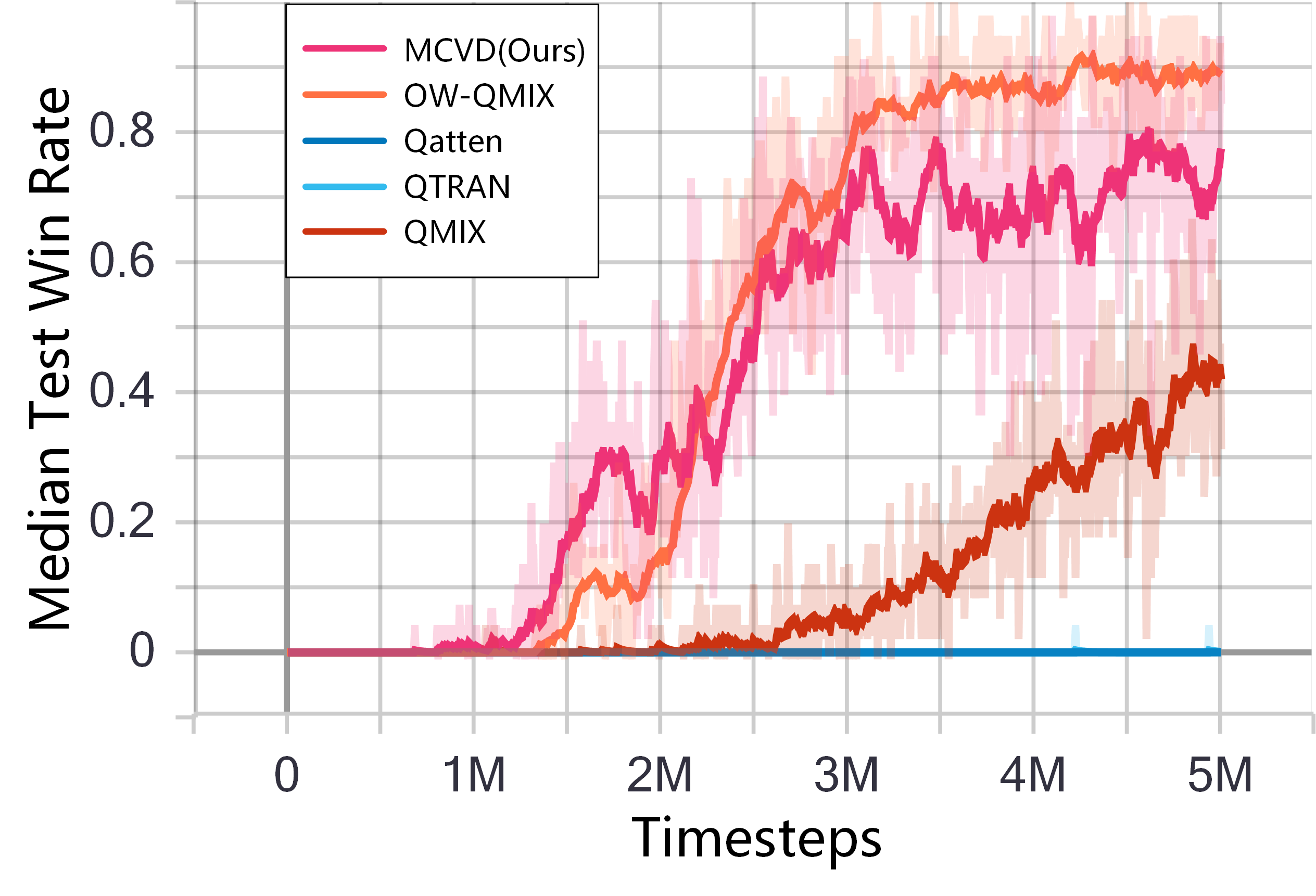}}
    \caption{Test win rate curves in multiple SMAC scenarios.}
    \label{fig:result-mcvd-smac}
\end{figure*}

As can be seen from Fig. 6, MCVD can achieve optimal or suboptimal in all scenarios, showing significant learning efficiency advantages in 2c\_vs\_64zg and 5m\_vs\_6m, only inferior to QMIX and OW-QMIX in 3s\_vs\_5z and MMM2, respectively, but also ranked suboptimal.
QTRAN and Qatten\cite{yang2020qatten} cannot always maintain stable learning performance, QMIX and OW-QMIX have occasional instability, while MCVD is the most stable one among all the algorithms.

This experiment demonstrates that MCVD can achieve comparable results even in environments where monotonic value decomposition algorithms excel. Based on the experiments in the Cooperative-Navigation scenario, a conclusion can be drawn that MCVD has a wide range of applicability for various experimental scenarios.

\section{Conclusion}
In this paper, we have revealed the irrationality of weight design in the existing Weighted QMIX operator and characterized the problem of value-decomposition as an Underfitting One-edged Robust Regression problem.
We have made the first attempt to give a solution to the value decomposition problem from the perspective of information-theoretical learning and introduced the Maximum Correntropy Criterion to the field of non-monotonic value decomposition. 
Through brief derivation, we have proposed a novel MARL algorithm called MCVD, which excels in the task of value decomposition in an unprecedentedly simple form.
Our experiment at OMG confirms that MCVD approaches the goal of $Q_{jt} \ge y$ as never before, both overestimation and underestimation are well eliminated.
Experiments in Cooperative-Navigation and multiple SMAC scenarios prove that MCVD has a wide range of applicability for MARL environments.
Compared with Weighted QMIX, it does not need parameter tunes according to specific environments and achieves the best stability among the compared algorithms.
The unprecedented ease of implementation, broad applicability, and stability of MCVD make it promising for widely used in engineering applications.


{\appendix[]
\section*{Ablation}
\label{Ablation}
In this section, we conducted ablation experiments in two simple SMAC scenarios, where the variables are value decomposition networks and joint action-value approximation networks.
\begin{figure}[]
    \centering
    \subfloat[2s3z]{\includegraphics[width = 2.3in]{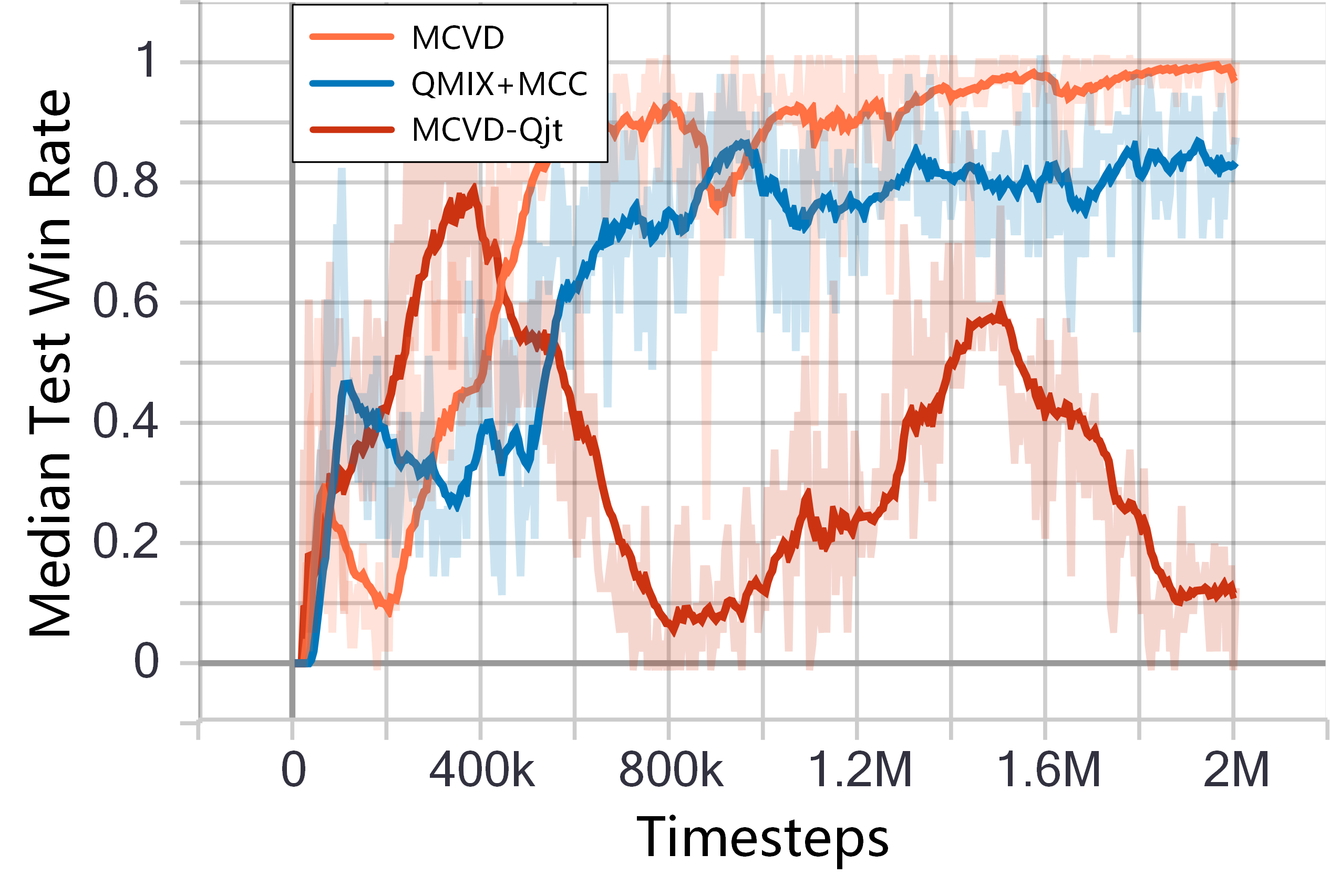}}\\
    \subfloat[3s5z]{\includegraphics[width = 2.3in]{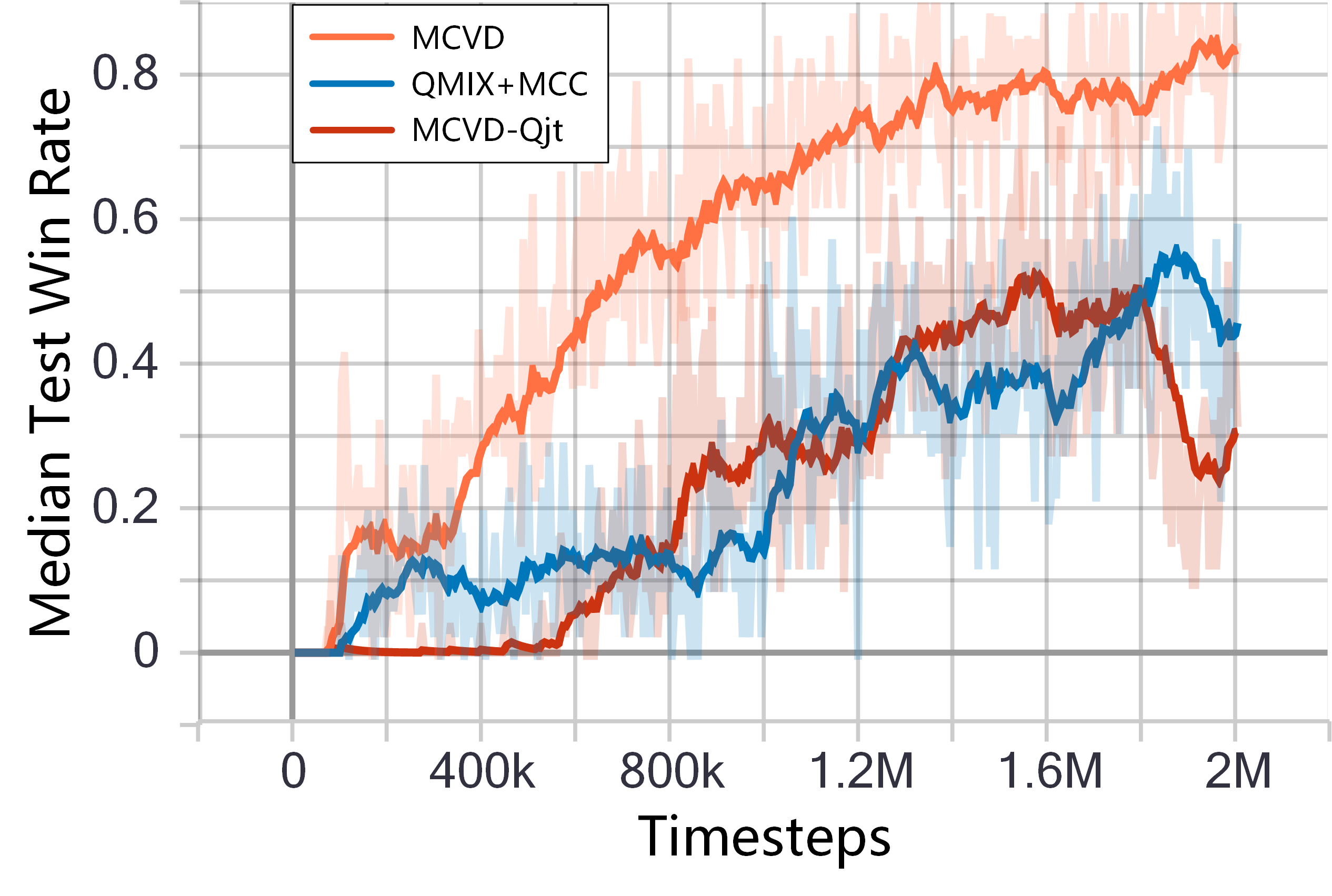}}
    \caption{Test return for two SMAC scenarios, comparing MCVD with two ablations.}
    \label{fig:ablation}
\end{figure}
where QMIX+MCC stands for replacing the summation network in MCVD with a weighted summation. 
That is, the value decomposition network evolves from VDN to QMIX.
MCVD-$Q_{jt}$ stands for not using the joint action-value approximation network.
It can be seen from Fig. \ref{fig:ablation} that the results of both ablations are poor, which supports our arguments in Sec. \ref{subsec details mcvd}.
\section*{Theoretical Analysis}
\label{appendix analysis}
Our theoretical analysis is based on the Theorem 1 and 2 of Weighted QMIX, please refer to the corresponding proofs in advance.
It should be noted that some symbols in this paper are different from that in Weighted QMIX, they are:
\[\begin{array}{l}
{Q_{jt}}(s,\bm{a}) \Leftrightarrow {Q_{total}}(s,\bm{u}),\\
\bm{\bar a} \Leftrightarrow {\bm{u^ * }},\\
{{\hat Q}_{jt}}(s,\bm{a}) \Leftrightarrow Q(s,\bm{u}),
\end{array}\]
on the left is the expression in this paper, and on the right is the expression in Weighted QMIX.
\subsection*{Impracticability of OW-QMIX}
According to the derivation of Weighted-QMIX, it can be known that the selection of weights needs to satisfy:
\begin{equation}
\begin{array}{l}
\alpha  = {\min _s}{\alpha _s} > 0,\\
0 < {\alpha _s} < \frac{{\Delta _s^2{{(1 - \gamma )}^2}}}{{{{({R_{\max }})}^2}{{\left| A \right|}^N}}},
\end{array}
\label{weight select theory}
\end{equation}
where ${\Delta _s} = {{\hat Q}_{jt}}(s,\bm{\bar a}) - \max \{ {{\hat Q}_{jt}}(s,\bm{a})|{{\hat Q}_{jt}}(s,\bm{a}) < {{\hat Q}_{jt}}(s,\bm{\bar a})\} $ to be the difference between the maximum joint action value and the second biggest joint action value\cite{bellemare2016gap}.
${R_{\max }} = \max r - \min r$, ${{{\left| A \right|}^n}}$ is the total of joint actions.

In general, ${\Delta _s^2{{(1 - \gamma )}^2}}$ can be treated as a constant, thus ${\alpha _s}$ should decrease squarely as ${R_{\max }}$ increases, and decrease exponentially as the number of agents increases. ${{{\left| A \right|}^N}}$ is known, but ${R_{\max }}$ is unknown, so the derivation can only stay at the theoretical level.
Since the theoretical value of ${\alpha _s}$ is very small in large-scale problems, and it is often accompanied by overestimation problems, thus greatly weakening the practicality of Weighted QMIX.


\subsection*{Rationality of MCVD}
Similar to Weighted QMIX, we derive the constraint on the kernel bandwidth $\sigma $ and explore its relationship with ${R_{\max }}$ and ${{{\left| A \right|}^N}}$.

We define ${\Pi _{mcc}}Q$ as the projection operator of MCVD, and let $\bm{\hat a} \in \arg \max {\Pi _{mcc}}Q$ denote the optimal combination of individual actions, and consider the loss when $\bm{\hat a} = \bm{\bar a}$ and $\bm{ a} \ne \bm{\bar a}$:
\begin{equation}
\begin{array}{l}
\mathop \mathbb{E}\limits_{\bm{a} \ne \bm{\bar a}} {{w_{td}}{{({{\hat Q}_{jt}}(s,\bm{a}) - {Q_{jt}}(s,\bm{a}))}^2}} \\
 = \mathop \mathbb{E}\limits_{\bm{a} \ne \bm{\bar a}} {\exp (-\frac{{{{({{\hat Q}_{jt}}(s,\bm{a}) - {Q_{jt}}(s,\bm{a}))}^2}}}{{2{\sigma ^2}}}){{({{\hat Q}_{jt}}(s,\bm{a}) - {Q_{jt}}(s,\bm{a}))}^2}}
\end{array}
\end{equation}

If we denote $e_r = {{\hat Q}_{jt}}(s,\bm{a}) - {Q_{jt}}(s,\bm{a})$, then the extreme point of ${\exp ( - \frac{{{e_r^2}}}{{2{\sigma ^2}}}){e_r^2}}$ is $e_r = \{  - \sqrt 2 \sigma ,0,\sqrt 2 \sigma \}$, where 0 is the minimum point, $\pm \sqrt 2 \sigma$ is the maximum point. Then the following inequality holds:
\begin{equation}
\begin{array}{l}
\mathop \mathbb{E}\limits_{\bm{a} \ne \bm{\bar a}} {\exp ( - \frac{{{e_r^2}}}{{2{\sigma ^2}}}){e_r^2}}\\
\le \mathop \mathbb{E}\limits_{\bm{a} \ne \bm{\bar a}} {\exp ( - \frac{{{{(\sqrt 2 \sigma )}^2}}}{{2{\sigma ^2}}}){{(\sqrt 2 \sigma )}^2}}\\
< {\frac{{2\sigma ^2}}{e}}{\left| A \right|^N}\\
\Rightarrow \mathop \mathbb{E}\limits_{\bm{a} \ne \bm{\bar a}} {{w_{td}}{{({{\hat Q}_{jt}}(s,\bm{a}) - {Q_{jt}}(s,\bm{a}))}^2}}  < {\frac{{2\sigma ^2}}{e}}{\left| A \right|^N}
\end{array}
\end{equation}
where $e$ is the nature constant.

And consider the loss when $\bm{\hat a} \ne \bm{\bar a}$ and $\bm{a} = \bm{\bar a}$, the derivation is independent of the weights, and the result is the same as Weighted QMIX:
\begin{equation}
\mathop \mathbb{E}\limits_{\bm{a} = \bm{\bar a}} {{w_{td}}{{({{\hat Q}_{jt}}(s,\bm{a}) - {Q_{jt}}(s,\bm{a}))}^2}}  \ge \Delta _s^2
\end{equation}

Let the loss of any ${{Q_{jt}}(s,\bm{a})}$ with $\bm{\hat a} \ne \bm{\bar a}$ and $\bm{a} = \bm{\bar a}$ greater than the loss of ${{Q_{jt}}(s,\bm{a})}$ with $\bm{\hat a} = \bm{\bar a}$ and $\bm{ a} \ne \bm{\bar a}$, thus the IGM could be achieved, we have:
\begin{equation}
\begin{array}{l}
{\frac{{2\sigma ^2}}{e}}{\left| A \right|^N} < \Delta _s^2\\
 \Rightarrow {\sigma ^2} < \frac{{e\Delta _s^2}}{{2{{\left| A \right|}^N}}}, \sigma  > 0,{\Delta _s} > 0\\
 \Rightarrow \sigma  < \sqrt{\frac{e}{2{{\left| A \right|}^N }}}\Delta _s
\end{array}
\end{equation}

It can be seen that with the increase of ${{{\left| A \right|}^N}}$, the kernel bandwidth $\sigma$ decreases much more slowly than the weight $\alpha$ due to the effect of the quadratic root, and it is not related with $R_{\max }$, which explains that a small kernel bandwidth is sufficient for most non-monotonic value decomposition problems. 
Due to the nature of MCC, a small kernel bandwidth does not seriously affect the gradient of small errors, thus the overestimation problem can be effectively avoided.
\section*{Experimental Setup}
\label{experimental setup}
\subsection*{Cooperative-Navigation with Discrete Action Space}
Cooperative-Navigation is a scenario of MPE where $N$ agents need to cooperate through physical actions to cover $L$ target landmarks in the shortest time.
An agent can observe their own positions and the relative positions of other agents and landmarks.
In this paper, the number of agents $N$ is set equal to the number of landmarks $L$.
The reward depends on the relative positions of agents and landmarks, and whether a collision occurs.

Cooperative-Navigation was initially proposed to verify MARL with continuous action space. There are no state and executable actions in the environment settings, and many interfaces are different from SMAC.
In order to achieve generalization and facilitate algorithm verification, we have improved Cooperative-Navigation so that its interface is almost consistent with SMAC.
At the same time, we averaged the original reward function to uniform the return curves with different numbers of agents.
\begin{figure}[H]
    \centering
    \includegraphics[width = 1.5in]{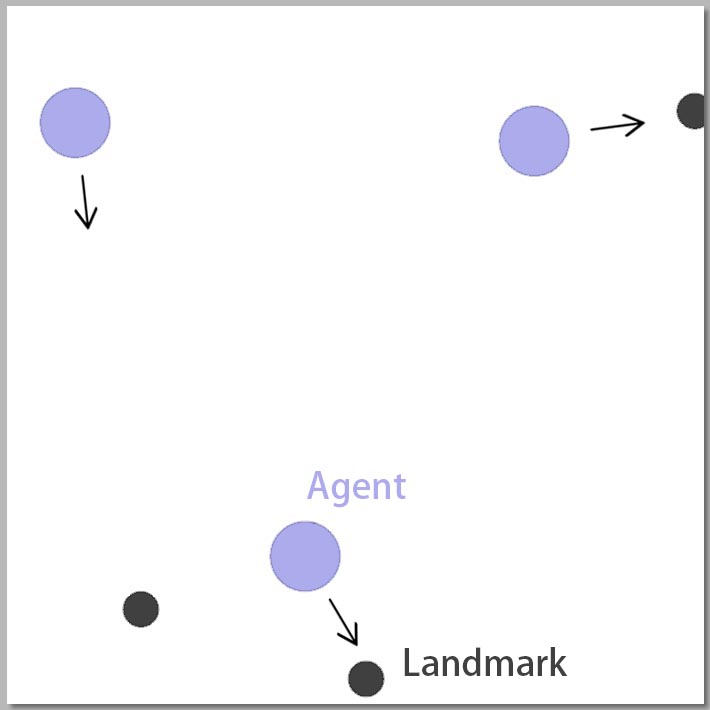}
    \caption{Illustration of Cooperative-Navigation scenario.}
    \label{fig:cp}
\end{figure}

The action dimension of Cooperative-Navigation scenarios is 5, which represents the acceleration directions of particle motion, which are [1: still, 2: upward, 3: downward, 4: left, 5: right].
Based on this, we can verify the non-monotonicity of the rewards through a simple and discrete scene.

We consider simplifying the problem as a grid world with two agents, assuming that the agents' velocity begins with 0, and the agents' position at the next moment is dependent on the direction of acceleration. The relative positions of the agent and the target point are shown in Tab. 4.
If the agent closest to a certain target point moves further, the reward of -1 will be accumulated, if closer, the reward of +1 will be accumulated, if it is stationary, the reward will be 0, and if a collision occurs, the reward of -10 will be accumulated, and the agents will go back to the last position, i.e. $s'=s$. The group reward is the cumulative value of the above rewards.
We just focus on the rewards when agent $\cal A$ takes actions 3 and 4, and agent $\cal B$ takes actions 1 and 4. The corresponding payoff is shown in Tab. 5.

\begin{table}[H]
    \centering
    \caption{Grid world expression of Cooperative-Navigation scenario.}
    \label{tab:cp scene}
    \begin{tabular}{|c|c|c|c|}
    \hline
        G1 & A& \\
        \hline
        &B&G2\\
        \hline
    \end{tabular}
    
\end{table}
\begin{table}[H]
\caption{Part of the payoff matrix of Matrix. \ref{tab:cp scene}}
    \label{tab:payoff cp}
    \centering
    \begin{tabular}{|c|c|c|}
    \hline
        \diagbox{$\cal A$}{$\cal B$} & 1&4 \\
        \hline
        3 &-10&0\\
        \hline
        4&1&0\\
        \hline
    \end{tabular}
    
\end{table}

According to the property of monotonicity, if ${Q_{{\cal B}}}(s \to {o_{{\cal B}}},1) > {Q_{{\cal B}}}(s \to {o_{{\cal B}}},4)$ holds, then should be ${Q_{jt}}({{\cal B}}:1,{{\cal A}}: \cdot ) > {Q_{jt}}({{\cal B}}:4,{{\cal A}}: \cdot )$, thus ${{\hat Q}_{jt}}({{\cal B}}:1,{{\cal A}}:3) < {{\hat Q}_{jt}}({{\cal B}}:4,{{\cal A}}:3)$ will become a contradiction. 
In contrast, if ${Q_{{\cal B}}}(s \to {o_{{\cal B}}},1) < {Q_{{\cal B}}}(s \to {o_{{\cal B}}},4)$ holds, then should be ${Q_{jt}}({{\cal B}}:1,{{\cal A}}: \cdot ) < {Q_{jt}}({{\cal B}}:4,{{\cal A}}: \cdot )$, thus ${{\hat Q}_{jt}}({{\cal B}}:1,{{\cal A}}:4) > {{\hat Q}_{jt}}({{\cal B}}:4,{{\cal A}}:4)$ will become a contradiction.

We suppose the non-monotonicity of this scenario mainly stems from the negative rewards brought by collisions. 
When the number of agents is large, the frequency of collisions will increase, and the non-monotonicity of the environment will also increase.

\subsection*{Platform}
All of the experiments are performed on the same server and under the same environment configuration. We use two HP power edge R740 workstations, each with two Intel Xeon Gold 6248 CPUs and a Tesla V100 GPU. The operating system is Ubuntu 20.04 and the environment is CUDA 11.6. All of the experiments are performed under Pytorch 1.7.1.

\subsection*{Hyperparameters}
See Table. \ref{tab:hyperparam}
\begin{table}[]
    \caption{Hyperparameters used in this paper.}
    \label{tab:hyperparam}
    \centering
    \begin{tabular}{c|c|c}
        \diagbox{Param}{Env} & SMAC&C-N\\
        \midrule
        difficulty & 7&None\\
        seed&123&123\\
        step\_mul&8&None\\
        n\_steps&2,000,000&500,000\\
        -&(MMM2: 5,000,000)&-\\
        train\_fre&1&1\\
        last\_action&True&True\\
        reuse\_network&True&True\\
        gamma&0.99&0.9\\
        optimizer&RMS&RMS\\
        evaluate\_fre&5000&5000\\
        evaluate\_epoch&32&32\\
        hidden\_dim&64&64\\
        lr&5e-4&5e-4\\
        batch\_size&32&32\\
        buffer\_size&5e3&5e3\\
        max\_epsilon&1&1\\
        min\_epsilon & 0.05&0.05\\
        anneal\_steps&50,000&50,000\\
        target\_update\_cycle&200&200\\
        grad\_norm\_clip&10&10\\
        $w$ for OW-QMIX&0.5&0.1\\
        $\sigma$ for MCVD&1&1\\
    \end{tabular}
\end{table}
}
%
\bibliographystyle{IEEEtran}
\bibliography{IEEEabrv,references_output}  

 





\end{document}